% This file contains a set of TeX macros for making manuscripts
%  in the style of astrophysical journals.
% Last modified Aug 29, 1990 to reflect the July 1, 1990 announcement
%  of style changes for most astronomical journals.
% Written by E. L. Robinson
%
% SUMMARY OF JRNLMACS MACROS
%
% \heading {Heading}
% \section {Section tTitle}
% \subsection {Subsection Title}
% \subsubsection {Subsubsection}
% \etal
% \eg
% \sun    % This puts in a subscripted circle with a dot.
% \journal   {authors} {year} {journal} {volume} {page}
% \infuture  {authors} {year} {journal} {a phrase}
% \book      {authors} {year} {title}          {pub. place}{publisher}
% \inbook    {authors} {year} {title} {editors}{pub. place}{publisher}{page}
% \infutbook {authors} {year} {title} {editors}{pub. place}{publisher}{a phrase}
% \privcom   {authors} {year} {a phrase}
% \phdthesis {author}  {year} {university}
% \circular  {authors} {year} {circular name} {a phrase like No. 2451}
% \samename
% \figcap   {number} {caption}
% \note {text}             % Creates a note at the bottom of the page.
% \apjnote {text}          % Creates a note in the middle of the page.
%
%
% First set up all the fonts.  This is necessary because
%  plainTeX uses too small a font size (10pt).
%  The default font size is changed to 12pt.
%  Note that the super- and sub-scripts have the same size
%  as the main text.  Although this does not look good, it
%  is Ap. J. style for manuscripts.  I rebelled at the prospect
%  of same sized scriptscrpts, however, and made them smaller.
\font\twelverm = cmr10 scaled\magstep1 \font\tenrm = cmr10
       
\font\twelvei = cmmi10 scaled\magstep1 
       \font\teni = cmmi10 
\font\twelveit = cmti10 scaled\magstep1 
       
\font\twelvesy = cmsy10 scaled\magstep1 
       \font\tensy = cmsy10 
\font\twelvebf = cmbx10 scaled\magstep1 \font\tenbf = cmbx10
       
\font\twelvesl = cmsl10 scaled\magstep1
\font\twelveit = cmti10 scaled\magstep1
\font\twelvett = cmtt10 scaled\magstep1
%
% Family 0 fonts are for ordinary text.
% Family 1 fonts are for math italics.
% Family 2 fonts are for ordinary math symbols.  They
%   must be the cmsy fonts.
% The special families are for text italic, slant roman, bold
%   face roman and typewriter fonts.
\textfont0 = \twelverm \scriptfont0 = \twelverm 
       \scriptscriptfont0 = \tenrm
       \def\rm{\fam0 \twelverm}
\textfont1 = \twelvei \scriptfont1 = \twelvei 
       \scriptscriptfont1 = \teni
       
\textfont2 = \twelvesy \scriptfont2 = \twelvesy 
       \scriptscriptfont2 = \tensy
       
\newfam\itfam \def\it{\fam\itfam \twelveit} \textfont\itfam=\twelveit
\newfam\slfam  \textfont\slfam=\twelvesl
\newfam\bffam \def\bf{\fam\bffam \twelvebf} \textfont\bffam=\twelvebf
       \scriptfont\bffam=\twelvebf \scriptscriptfont\bffam=\tenbf
\newfam\ttfam  \textfont\ttfam=\twelvett
\rm
\hsize=6in
\hoffset=.45in
\vsize=9in
\baselineskip=24pt
%
% The Ap. J. does not like hyphens or right justification.  Okay,
%  we try to oblige -- within limits!
\raggedright  \pretolerance = 800  \tolerance = 1100
\raggedbottom
%
% The following two lines set up the default dimensions to be 
%  used for headings of sections and subsections.
\dimen1=\baselineskip \multiply\dimen1 by 3 \divide\dimen1 by 4
\dimen2=\dimen1 \divide\dimen2 by 2
%
% The following macro puts the page numbers at the top of the page
%  and suppresses the page number on the title page.
\nopagenumbers
\headline={\ifnum\pageno=1 \hss\thinspace\hss 
     \else\hss\folio\hss \fi}
%
% The following macro makes unnumbered headings.  They are useful
% for the abstract page, for table headings, and so on.
 
%
% The following macros make sections numbered with arabic numerals.
%  The macro keeps track of section numbers automatically.
\newcount\sectcount
\newcount\subcount
\newcount\subsubcount
\global\sectcount=0
\global\subcount=0
\global\subsubcount=0
\def\section#1{\vfill\eject \vbox to \dimen1 {\vfill}
    \global\advance\sectcount by 1
    \centerline{\bf \the\sectcount.\ \ {#1}}
    \global\subcount=0  %  This definition resets the subsection number.
    \global\subsubcount=0  %  This definition resets the sub-subsection number.
    \vskip \dimen1}
%
% The following macro makes subsections like 3.1, 3.2 and so on.
% The macro keeps track of subsection numbers automatically.
\def\subsection#1{\global\advance\subcount by 1 
    \vskip \parskip  \vskip \dimen2
    \centerline{{\it \the\sectcount.\the\subcount.\ \ #1}} 
    \global\subsubcount=0  %  This definition resets the sub-subsection number.
    \vskip \dimen2}
%
% The following macro makes sub-subsections like 3.1.1, 3.1.2 and so on.
% The macro keeps track of sub-subsection numbers automatically.
\def\subsubsection#1{\global\advance\subsubcount by 1 
    \vskip \parskip  \vskip \dimen2
    \centerline{{\it \the\sectcount.\the\subcount.\the\subsubcount.\ \ #1}} 
    \vskip \dimen2}
%
% The following are reference templates. Note the preliminary
% definition.
%
\def\refindent{\advance\leftskip by 24pt \parindent=-24pt}
%
% This template is useful for referencing articles in journals.

%
% And this one is useful if the article is submitted or in press.

%
% This template is useful for referencing articles in books.

%
% This template is useful for referencing articles in  that
% are scheduled to appear in a book that will be published in the future.

%
% This template is useful for referencing books.

%
% This template is for private communications and preprints going
%  to unknown journals.

%
% This template is for referencing Ph.D. theses.

%
% This template is for referncing circulars and bulletins that
%  do not have a volume number, such as IAU Circulars or IAU
%  Inf. Bull. Var. Stars.

%
% The figcap macro is can be used to set up the figure captions.

%
% And now some useful definitions
\def\etal{{\it et al.\/\ }}
\def\eg{{\it e.g.\/}}
\def\sun{_\odot}

%
%  Here are two footnote macros.
%
%  The \note{} macro creates numbered footnotes for use on the
%  titlepage of the manuscript.  Because the footnotes are on the
%  titlepage, they are put at the bottom of the page.
%  \note automatically numbers the footnotes.
%  Usage:  \note{text}
%  Note that the value of \footstrut has been changed to give proper 
%  spacing between footnotes.
%
%  The \apjnote macro creates numbered footnotes for use in the
%  text of the manuscript.  These footnotes are inserted in the
%  middle of the page, not at the bottom.
%  \apjnote automatically numbers footnotes, the numbers continuing
%  correctly even if \note has been invoked on the titlepage.
%  Usage: \apjnote{text}

\newcount\notenumber
\notenumber=0
\def\note{\global\advance\notenumber by 1 
	  \footnote{$^{\the\notenumber}$}}
\def\apjnote#1{\global\advance \notenumber by 1 
	       $^{\the\notenumber}$ \kern -.6em
               \vadjust{\midinsert
                        \hbox to \hsize{\hrulefill}
                        $^{\the\notenumber}${#1} \hfill \break
                        \hbox to \hsize{\hrulefill}
                        \endinsert
                       }
              }
{\hoffset = 0.25in
\vbox to 0.7in{\vfill}
\vskip .5in
\centerline{\bf HIGH SPATIAL RESOLUTION KAO FAR-INFRARED OBSERVATIONS}
\vskip .1in
\centerline{\bf OF THE CENTRAL REGIONS OF INFRARED-BRIGHT GALAXIES}
\vskip .1in
\vskip .2in
\centerline{Beverly J. Smith$^1$ and
P. M. Harvey}
\centerline{Department of Astronomy}
\centerline{The University of Texas at Austin}
\centerline{Austin TX 78712}
\vskip 1.0in
$^1$Now at IPAC/Caltech, MS 100-22, Pasadena CA  91125.
\vskip 1.0in
\centerline{Received:{\hbox to 1.8in {\thinspace\hrulefill}};
Accepted:{\hbox to 1.8in {\thinspace\hrulefill}}}
\vskip 1.0in
\vfill \eject

\def\degree#1{\ifmmode{\if.#1{{^\circ}\llap.}\else{^\circ} #1\fi}\else
{\if.#1$^\circ$\llap.\else\if\empty#1$^\circ$#1\else$^\circ$ #1\fi\fi}\fi}
\def\arcmin#1{\ifmmode{\if.#1{'\llap.}\else{'} #1\fi}\else
{\if.#1$'$\llap.\else$'$ #1\fi}\fi}
\def\arcsec#1{\ifmmode{\if.#1{''\llap.}\else{''} #1\fi}\else
{\if.#1$''$\llap.\else$''$ #1\fi}\fi}
\def\arcss#1{\ifmmode{\if.#1{\rm s\llap.}\else{\rm s} #1\fi}\else
{\if.#1${\rm s}$\llap.\else${\rm s}$ #1\fi}\fi}
%-----------------------------------------------------------------
\def\ref{\par\noindent \hangindent=3em \hangafter=1}

%-----------------------------------------------------------------
\def\item {{\medskip \noindent $\bullet{   }$ }}
\def\jref#1 #2 #3 #4 {{\par\noindent \hangindent=3em \hangafter=1 
      #1, {\it#2}, {\bf#3}, #4.\par}}
\def\ref#1{{\par\noindent \hangindent=3em \hangafter=1 
      #1.\par}}
\def\Ha{\ifmmode H\alpha\else H$\alpha$\fi}
\def\LIR{\ifmmode L_{IR}\else L$_{IR}$\fi}
\def\LCO{\ifmmode L_{CO}\else L$_{CO}$\fi}
\def\ICO{\ifmmode I_{CO}\else I$_{CO}$\fi}

\def\I.{\kern.2em{\sc i}}  \def\II.{\kern.2em{\sc ii}} 
\def\eg.{{\it e.g.}}
\def\ie.{{\it i.e.}}
\def\etal.{{\it et al.}}
%-----------------------------------------------------------------

{\bf ABSTRACT}

We present new high spatial resolution Kuiper Airborne Observatory
(KAO) 50 $\mu$m and/or
100 $\mu$m data for
11 infrared-bright galaxies.
The sizes 
of the central far-infrared emitting
regions in three of these galaxies, NGC 660, NGC 891,
and M 83, agree 
with those
of the central 
star formation complexes.
The Sb galaxy NGC 7331, which does not
have observed star formation in its bulge,
has a flat-topped
100 $\mu$m major axis profile which covers
the bulge and
inner spiral arms, without a bright central peak or
pronounced depression.
The remaining seven galaxies, all of which
are known to have strong nuclear or
circumnuclear star formation, are unresolved
or marginally resolved with the KAO, with far-infrared
source size limits 
consistent with the sizes of the central star formation
complexes.

Along with these new data, we have tabulated previously
published KAO data for
11 other galaxies, and IRAS
60 and 100 $\mu$m data for the bulges of the
large angular size galaxies M 31 and M 81.  
From the literature,
we have compiled
optical, near-infrared, and millimeter measurements for the
central regions of
the entire set of 24 galaxies.
We have used this dataset to investigate dust heating and 
star formation in the central areas of galaxies.
We find
that L(FIR)/L(B) and L(FIR)/L(H) correlate
with 
CO (1 $-$ 0) intensity and 100 $\mu$m
optical depth.
Galaxies with optical or near-infrared signatures of 
OB star formation 
in their central regions have higher values of I(CO) and $\tau$$_{100}$
than more quiescent galaxies,
as well as higher far-infrared surface brightnesses and 
L(FIR)/L(B) and L(FIR)/L(H) ratios.
The L(FIR)/L(H$\alpha$) ratio does not correlate
strongly with CO and $\tau$$_{100}$.
These data are consistent with a scenario in which OB
stars dominate dust heating in the more active
galaxies and older stars are important for the more
quiescent bulges.
Whether or not a galaxy bulge has strong star formation
may be decided by a threshold effect;
star forming galaxies have surface gas densities
above the Kennicutt critical density, while
quiescent galaxies have lower values.

{\bf 1. Introduction }

It has been known for more than 25 years that galaxies
emit a significant amount of
their total luminosity at wavelengths longer than 30 $\mu$m,
and that this radiation originates from dust in the interstellar
medium
(Harper and Low 1973; Telesco, Harper,
and Loewenstein 1976; Telesco and Harper 1980).
The successful IRAS mission in 1983 
(IRAS Explanatory Supplement 1988)
provided far-infrared luminosities of
thousands of galaxies, permitting many investigations into
the nature of this far-infrared emission.
Two such studies were those
of Helou (1986) and Persson and Helou (1987),
who suggested a two-component model for the far-infrared emission
from galaxies.  In this scenario, both OB stars and 
the general interstellar radiation field contribute to dust heating
in galaxies, in proportions differing from galaxy to galaxy.
In active galaxies, the AGN may also contribute to dust heating.

Since these publications, there have been many efforts to 
test and refine this picture.
One difficulty
has been the relative lack of
spatially resolved far-infrared
measurements of galaxies.
The IRAS resolution
is relatively poor (1\arcmin.5 $\times$ 5$'$ at 100 $\mu$m),
so without special processing, only the most nearby
galaxies can be studied in detail 
(Walterbos and Schwering
1987; Deul 1989; 
Schwering and Israel 1989; Schwering 1989;
Rice et al. 1990;
Gallagher et al. 1991).
The development of the HiRes deconvolution method for 
IRAS data (Aumann,
Fowler, and Melnyk 1990) has allowed studies of more
distant galaxies 
with an effective resolution of $\sim$80$''$ at 100 $\mu$m 
(Marston 1989; Rand, Kulkarni, and Rice 1992;
Devereux and Young 
1992, 1993; 
Rice 1993; 
Surace et al. 1993;
Devereux
and Scowen 1994;
Marsh and Helou 1995).

To obtain information at smaller physical scales
or in more distant galaxies
even higher resolution is needed.
To this end,
we have conducted a program of 
far-infrared observations of galaxies using 
the University
of Texas far-infrared photometer systems
on 
the Kuiper
Airborne Observatory (KAO).  These systems
have a 100 $\mu$m beamsize of approximately
30$''$ $\times$ 40$''$.
In a series of previous papers, we have presented 
KAO far-infrared data for
a number of 
galaxies (Joy et al. 1986, 1987,
1988, 1989;
Lester et al. 1986a, 1987, 1995;
Brock et al. 1988; 
Smith et al. 1991, 1994, 1995).
In these studies we have identified 
circumnuclear rings of far-infrared emission
(Lester et al. 1986a; Smith et al. 1991), 
resolved central star formation complexes
(Lester et al. 1987; Joy et al. 1987, 1988;
Smith et al. 1995), separated
bulge and spiral arm emission in
normal galaxies (Smith et al. 1991, 1994),
resolved the far-infrared emission from
a distant giant elliptical galaxy (Lester et al. 1995),
separated a close
galaxy pair (Joy et al. 1989),
and obtained stringent limits on the sizes of 
the far-infrared emitting regions in luminous galaxies
(Joy et al. 1986; Brock et al. 1988).
For a more detailed review of these publications,
see Lester (1995).

In the current paper, we present new KAO data for 11 additional
galaxies.
We combine these new data with previously-published
KAO data for 11 galaxies, along
with IRAS far-infrared data for the bulges of M 31 and M 81.
For the entire sample of 24 galaxies,
we compare the far-infrared data with published optical,
near-infrared, and millimeter observations to
investigate dust heating and star formation
in the bright central regions
of galaxies.

{\bf 2. The Observations }

The observed galaxies are given in Table 1, along with 
their positions,
distances, morphological and spectral types, blue
magnitudes, and optical diameters.
The far-infrared data were obtained with the
KAO 0.9m telescope during the period
1987 - 1994, flying from
either Moffett Field, California, Hickam Air Force Base, 
Hawaii,
or Christchurch, New Zealand.
Ten of the 11 galaxies were observed at 100 $\mu$m, and three
were observed at 50 $\mu$m.
Details on the observations of individual galaxies
are given in Table 2.

The observations were made with either 
a 8$\times$1 bolometer array (Low et al. 1987; Smith
et al. 1991)
or a 10$\times$2 element array (Smith et al. 1994).
For both these systems, the approximate rms noise level
reached in an hour is 0.5 Jy.
The 100 $\mu$m beamsizes of these systems are
approximately
25$''$ $\times$ 35$''$ (8 channel system) or
31$''$ $\times$ 41$''$ (20 channel system),
varying slightly from detector to detector, with
the short axis of the beam aligned along the long
axis of the array.  The 50 $\mu$m beamsize 
for both systems is approximately 18$''$ $\times$ 25$''$.
Along the long axis of the array, the detectors
are separated by approximately half a beamwidth (see
Smith et al. 1991, 1994),
while the two banks of detectors in the 20 channel
system are separated by 28$''$ at 100 $\mu$m and
18$''$ at 50 $\mu$m.

With the 8 channel system, the observations 
were made with a scanning mode (see Smith et al. 1991),
where the array was scanned across
the galaxy perpendicular to the array axis,
resulting in 8 simultaneous azimuthal scans across the galaxy.
The scan lengths ranged from 2\arcmin.3 to 3\arcmin.2, and
the data were binned at 2$''$ intervals with
a scan rate of 5$''$ sec$^{-1}$.
With the 20 channel system, the scanning
method was used for NGC 3256, giving
10 parallel scans across the galaxy, while a nodding mode 
(see Smith et al. 1994) was used for
four galaxies.
In this method, the
array was nodded on and off the galaxy at
10 second intervals.
For NGC 891, data were taken
at 2 overlapping array locations on the sky,
shifting along the major axis of the array.
For NGC 7331, three overlapping sets of 
data were obtained.
Chopping was in the azimuthal
direction, with amplitudes 
which varied from 3$'$ $-$ 6\arcmin.5.
For both of these systems, the long axis of the array
was aligned with the altitude axis of the telescope, 
causing the position angle of the array on
the sky to vary 
during a long integration.
The ranges in these rotation angles are given in Table 2.
With the exception of NGC 1808 and NGC 6946, these rotation
angles changed less than 12$^{\circ}$ during the observations.
For the NGC 7331 observations,
the KAO `freeze mode' was used,
keeping the rotation angle steady to $\le$1$^{\circ}$.

For the April 22, 1988 observations
of M~83 and for NGC~7331, we guided on the nucleus of the galaxy itself.
For the rest, offset guiding was used, since the nuclei
were too faint to guide on.
The combination of offset guiding and diurnal motion causes the
position of the center of the array on the sky to vary with time.
To compensate for this effect, the offset was re-calculated and
adjusted after each half sweep 
to keep the galaxy centered on the array.

To obtain a point source profile and to calibrate the data,
we also observed at least one of the
following objects during each flight:
IRC +10216, IRC +10420, $\eta$ Carinae,
Ceres, Vesta, Callisto, or Uranus.  
Details on these observations are also given in Table 2.
Callisto, the asteroids, and Uranus all have angular diameters less than
4$''$, so are clearly point sources.
IRC +10216, IRC +10420, and $\eta$ Carinae have also been shown to be point
sources at 50 and 100 $\mu$m at this resolution
(Lester, Harvey, and Joy 1986b; Harvey et al. 1991; Harvey,
Hoffmann, and Campbell 1978).

{\bf 3. The Data }

After noise spike removal, the scan data were offset-adjusted, filtered,
flat-fielded, and co-added as in Smith et al. (1991).
On the May 23, 1990 flight, strong correlated noise 
between the detectors was seen.
To eliminate this noise, for those observations
we 
subtracted the signal in channel 1 from that in the rest of
the detectors before co-addition.

The final summed scans 
are shown in Figure 1.
These plots show that the galaxies are all relatively compact
compared to the instrumental profile.
To make this comparison more directly,
in Figure 2 we plot the galaxy and corresponding
calibration data for
the channel with the strongest signal.
These plots show that 
M 83 is 
extended with respect to the point
source, while NGC 253 and NGC 6946 appear marginally
resolved. The rest
are unresolved.
To make this assessment more quantitative, we determined the
rms deviation of each galaxy profile from the point source profile,
within the range where the galaxy data remained above zero.
Only for M 83 is the rms greater than 3$\sigma$.
For the rest, we consider our derived source sizes to be upper limits.

To get an estimate of the source size in the cross-scan direction,
we averaged the central 22$''$ (100 $\mu$m)
or 10$''$ (50 $\mu$m) of the summed scans for each
detector. 
These profiles along the long axis of the array are
given in Figure 3.
None of the galaxies are strongly resolved 
relative to the point source profiles.
NGC 1097 and NGC 6946 appear marginally resolved,
with possible low-level extended
emission in the end channels, while the rest are unresolved.

The nod data were co-added as in
Smith et al. (1994), and the summed data profiles are shown in
Figure 4.
These plots show that, if the far-infrared
peak was well-centered on the first bank of detectors,
then NGC 660, M 83, and NGC 7331 are
resolved along the long axis of the array,
while NGC 891 is unresolved.
NGC 7331 is the most extended galaxy in this sample,
showing a very broad flat-topped profile with
an approximate FWHM size of 165$''$.
The nod data for M 83 (Figure 4c) show a resolved central source
with surrounding extended emission.
To test whether the central peak is
truly resolved above this plateau, we have re-plotted
these data in Figure 4d with the end channels
set to zero.  In this figure, the central source appears only
marginally extended, so we conservatively consider
our derived size an upper limit.

All four galaxies in Figure 4
appear resolved along the minor
axis of the array, 
if we are correctly centered with the infrared peak in the
first bank of detectors.
The flux observed in the second bank of detectors (lower
panels in Figure 4a-e) is higher for the galaxies
than for the point source profiles.  

In Table 3, we give an estimate of the position of the
far-infrared maxima in the sample galaxies.  For the
scan data, these are simply
the location of the peak position for the detector with
the strongest signal; we did not attempt to interpolate
between detectors.  
For the M 83 and NGC 660 nod data, the positions given
are the location of the detector with the strongest
signal, and for NGC 891, we have averaged the positions of
the two detectors in the first bank which have approximately
equal flux densities.
For NGC 7331, which does
not have a strong far-infrared peak, we give the
approximate center of the broad profile seen in
the data from the first bank.
Our absolute pointing accuracy is
10$''$.
For galaxies which are poorly
centered on the array, 
there is an additional uncertainty 
in the location of the far-infrared peak.
In Table 1, we provide the best available published
measurements
of the positions of the nuclei of these galaxies,
from either radio or optical data.
The far-infrared positions in Table 3 
are all consistent with these nuclear
positions within our uncertainties.
Therefore, for the nod data shown in
Figure 4, if the far-infrared distribution is
symmetrical along the minor axis of the array with respect to
the nucleus, then the galaxy is spatially resolved in
this direction.

From the data plotted in Figures 2, 3, and 4, we have made
estimates of the intrinsic sizes of the central
far-infrared sources in these galaxies
using a simple Gaussian deconvolution.
These results are given in Table 3.
The FWHM sizes range from
$\le$13$''$ to 165$''$ ($\le$220 pc $-$ 12 kpc).
For the nod data, the minor axis sizes were determined
assuming symmetry about the nucleus and so are
very uncertain.

The calibrated flux densities 
are given in Table 3.
We estimate that these measurements are accurate to 30$\%$.
For these calibrations, we assumed that
Uranus is a blackbody of
58.5K (Stier et al. 1978) with a diameter of 50,800 km, and Callisto
is a blackbody of temperature of 147K 
(Campbell et al. 1989) and diameter 5020 km (Jones and Morrison 1974).
For Ceres and Vesta, we used a standard thermal model of the far-infrared
emission from asteroids (Lebofsky et al. 1986), along
with the IRAS diameters of 848 km and 468 km and
albedos of 0.11 and 0.42, respectively (IRAS Minor
Planet Survey 1992).

Determining absolute flux densities using
the evolved stars $\eta$ Carinae, IRC +10420, and IRC +10216 
requires cross-calibration due to the possibility
of far-infrared variability.
To calibrate the March 1992 M 83 data, we use an $\eta$ Carinae 
100 $\mu$m
flux density of 5200 Jy (Harvey et al. 1978), since
we confirmed
earlier in
the same flight series
that the 100 $\mu$m flux density
had not changed
(Smith et al. 1994).
We cross-correlated the August 1990 IRC +10420
data using Hygeia observations from a different flight
during the same flight series,
and find values consistent with
the Harvey and Wilking (1982) 
flux densities of F(50) = 930 Jy and F(100) = 240 Jy
(Smith, Harvey, and Lester 1995).
Also, using the August 23, 1990 Vesta data to
calibrate IRC +10420 gives a consistent value of F(50)
= 941 Jy.
For the August 1994 observations
we have no independent calibration
of IRC +10420, however, observations
of Uranus during the same flight
series give a calibration
conversion consistent with that obtained using
IRC +10420 with the flux densities given in Harvey
and Wilking (1982).
For the April 1988 M 83 data,
we used the PSC color-corrected 100 $\mu$m flux 
density of IRC +10216 of
862 Jy.
This agrees within 15$\%$ with
our independent calibration 
of 
F(100) = 759 Jy
for
IRC +10216
based on Uranus observations
using unpublished
April 1989
data.
Table 3 shows
that the peak M 83 brightnesses from the two different
observations agree well.

In Table 3, we provide both a peak flux density
and a total flux density.
In some cases, because the galaxy was not
well-centered on the array,
the peak flux density does not equal the total flux
density
although the galaxy is unresolved.  These incidences
are noted in Table 3.
In Table 3, we also give the total IRAS 
flux densities for the galaxies, obtained either from
Rice et al. (1988) or IRAS Addscans.
At 50 $\mu$m,
we estimate total flux densities from the
IRAS 25 and 60 $\mu$m data assuming a $\lambda$$^{-1}$
emissivity law. 
The total flux densities from the KAO agree with the total IRAS flux densities
within the
uncertainties for all of the galaxies
except for NGC 891, M~83, and NGC 6946.
For these galaxies, our KAO observations are clearly missing
extended emission.
For the rest of the galaxies,
less than 30$\%$ of the total far-infrared
emission arises outside of the region observed by the KAO.

Three of the galaxies in our sample were previously
observed 
at 50 or 100 $\mu$m in a small beam 
(Rickard and Harvey 1984;
Engargiola 1991).
Our new peak brightness measurements 
are consistent with these data within the uncertainties
(see Table 3).  

{\bf 4. Discussion of Individual Galaxies }

Comparing the far-infrared morphology of individual galaxies
with the observed distribution of star formation regions and
molecular gas is useful in determining the origin
of extragalactic far-infrared radiation.
In this section, we 
discuss 
each galaxy in our sample individually, summarizing
the available data at other wavelengths,
and noting how the far-infrared
distribution compares with that of the other components
of the galaxy.

{\it 4.1. NGC 253}

NGC 253 is a highly inclined spiral galaxy
with 
a weak bar (Pence 1980; Scoville et al. 1988).
Near-infrared spectroscopy of the nuclear region of
NGC 253 reveals an obscured central starburst
(Wynn-Williams et al. 1979; Rieke et al. 1980;
Beck and Beckwith 1984; Rieke, Lebofsky, and Walker 1988).
The CO (1 $-$ 0) and (2 $-$ 1) distributions are 
centrally peaked (Scoville et al. 1988; Mauersberger
et al. 1995), and the CO (2 $-$ 1) size
is  $\sim$40$''$ $\times$ 10$''$ FWHM elongated
east-west (Wall et al. 1991; Mauersberger et al. 1995).
Denser gas appears to be more centrally concentrated;
the $^{13}$CO (3 $-$ 2) source 
FWHM is $\le$15$''$ (Wall et al. 1991).
The 1.3 mm and 3 mm FWHM sizes are 16$''$ $\times$
11$''$ and 10$''$ $\times$ 4$''$, respectively (Kr\"ugel
et al. 1990; Carlstrom et al. 1990).
At 10 $-$ 30 $\mu$m, the FWHM size is 7$''$ $\times$ $\le$4$''$
(Telesco, Dressel, and Wolstencroft 1993).
Our far-infrared source size limits ($\le$16$''$
$\times$ 14$''$
and  
$\le$21$''$ $\times$ $\le$31$''$ at 50 and 100 $\mu$m, respectively), 
show that the far-infrared emitting region, particularly
at 50 $\mu$m, is more compact that the CO (2 $-$ 1)
emission,
but consistent with the sizes at the other wavelengths.

Low-level extended far-infrared disk emission is observed
in NGC 253 
out
a diameter
of 26\arcmin.5 at 60 $\mu$m (Rice et al. 1988).
Our KAO observations show that this extended emission is
a relatively small percentage, $\le$30$\%$, of the
total far-infrared emission from the galaxy, with the 
majority of the radiation arising from the central unresolved source.

{\it 4.2. NGC 660}

NGC 660 is a peculiar barred spiral galaxy with a possible
polar ring (Young, Kleinmann, and Allen 1988).
It has an `N'-type optical spectrum (Keel 1984),
however, H$\alpha$+[N~II] mapping indicates significant
massive star formation over an extended region 
in the center of the galaxy (Young
et al. 1988).  The CO distribution is extended northwest
to southeast along the bar, with a FWHM size of $\sim$22$''$
$\times$ 35$''$ (van Driel et al. 1995). 
We find a FWHM size of the 50 $\mu$m emitting region
of 22$''$ $\times$ $\sim$46$''$ (1.0 kpc
$\times$ 2.1 kpc), similar to the CO size.

{\it 4.3. NGC 891}

NGC 891 is an edge-on Sb galaxy 
thought to be very similar to the Milky Way.
The CO (1 $-$ 0) distribution is centrally peaked with
a surrounding molecular ring (Scoville et al. 1993),
while the HI map shows a central depression (Rupen 1991).
In the IRAS data, NGC 891 was resolved along its
major axis but unresolved along its minor axis 
(Wainscoat, de Jong, and Wesselius 1987). 
Along the major axis, the far-infrared 
traces the radio continuum
and CO (1 $-$ 0) emission well (Wainscoat et al.
1987).  
With the KAO, we have further constrained the
100 $\mu$m size of NGC 891 along its
minor axis to be $\le$37$''$.
NGC 891 is extended along its major axis in the far-infrared,
with $\sim$25$\%$ of the total 100 $\mu$m emission
coming from the central
31$''$ $\times$ 41$''$.

{\it 4.4. NGC 1097}

NGC 1097 is a beautiful barred spiral with a 
LINER nucleus surrounded by a ring of star formation of
diameter 15$''$ $-$ 20$''$ (Osmer, Smith, and Weedman 1974;
Meaburn et al. 1981).
This ring is present in radio continuum and H$\alpha$ maps
(Hummel et al. 1987), at 10 $\mu$m (Telesco and Gatley 1981;
Telesco et al. 1993),
and in the CO (J = 1 -- 0) line (G\'erin et al. 1987).
Our 100 $\mu$m data show a central unresolved core
with a source size of $\le$36$''$ $\times$ $\le$41$''$, consistent
with the size of the star forming ring.  

{\it 4.5. NGC 1808}

The circumnuclear region of NGC 1808 contains several
blue `hot spot' H~II regions (Morgan 1958; S\'ersic and
Pastoriza 1965; V\'eron-Cetty and V\'eron 1985;
Phillips 1993).
The optical
spectrum of the nucleus itself was classified as that
of a Seyfert galaxy
by V\'eron-Cetty and V\'eron (1985), but as indicative
of OB stars by Phillips (1993).
The central star formation complex has a size of 10$''$
$\times$ 20$''$, extended southeast to northwest (V\'eron-Cetty
and V\'eron 1985).
Outflow from this central area is indicated both
by peculiar dust filaments extending out of this
region perpendicular to the plane of the galaxy
(Burbidge and Burbidge 1968;
V\'eron-Cetty and V\'eron 1985) and by velocity measurements
of the 
ionized gas (Phillips 1993).
Our 100 $\mu$m source size limit of $\le$29$''$ $\times$ $\le$34$''$
is consistent with that of this star forming region.

{\it 4.6. NGC 3256}

NGC 3256, a peculiar galaxy with two long tidal tails (Sandage
and Brucato 1979; Joseph and Wright 1985),
is likely the remnant of the merger between two spiral galaxies
(Graham et al. 1984).
Ultraviolet, optical, 
and near-infrared spectroscopy of this galaxy indicate that it
is undergoing a strong central starburst (Kinney et al.
1993; Storchi-Bergmann, Kinney, and Challis 1995;
Kawara, Nishida, and Gregory 1987;
Doyon, Joseph, and Wright 1994).
Br$\gamma$ emission is detected over
over a 27$''$ $\times$ 16$''$ region, with a FWHM
size of 8$''$ $\times$ 5$''$ (Moorwood and Oliva 1994a).
Emission at 
10 $\mu$m is seen to a radius of 10$''$, with
a FWHM size of 9$''$ $\times$ 7$''$ (Graham et al. 1987).
NGC 3256 has been mapped in the CO (2 $-$ 1) line by
several groups (Sargent, Sanders, and Phillips 1989;
Casoli et al. 1991; Aalto et al. 1991).
Casoli et al. (1991) determine an intrinsic FWHM size
of 25$''$ $\times$ 11$''$, while Aalto et al. (1991)
find a smaller FWHM size of
11$''$ $\times$ 6$''$.
Our far-infrared source size limit of
$\le$24$''$ $\times$ $\le$22$''$ is consistent with these
measurements.

{\it 4.7. M 83}

The well-known barred spiral galaxy M 83 is undergoing rapid star formation
in and around the nucleus, at the ends of the bar,
and on the spiral arms 
(Talbot, Jensen, and Dufour 1979; Ondrechen 1985; Bohlin et al.
1983; Gallais et al. 1991).
The HI
distribution has a 2$'$ diameter central depression  
(Allen, Atherton, and Tilanus 1986), 
while the CO (1 $-$ 0) distribution is centrally peaked and traces
the optical bar (Handa et al. 1990).

The IRAS data on M 83 show that it is quite extended in the far-infrared;
Rice et al. (1988) give a 60 $\mu$m 
size
(to 25 mJy arcmin$^{-2}$)
of 15.6$'$.
Marston (1989) deconvolved IRAS pointed observations data on M 83,
and revealed a possible far-infrared bar.
Our KAO data resolves the central 100 $\mu$m source,
giving a FWHM of 20$''$ $\times$ $\le$26$''$.
Thus the far-infrared emission from this central source 
has a size consistent with that 
of the central CO source
(Handa et al. 1990).
Approximately 20$\%$ of the total 100 $\mu$m emission from
this galaxy arises from this central source.

{\it 4.8. The Circinus Galaxy}

The Circinus galaxy (Freeman et al. 1977) 
is a large, nearby spiral galaxy with 
a compact nuclear radio source, optical and infrared
spectra
indicative of non-thermal activity, and
nuclear H$_2$O maser emission (Moorwood and Glass 1984;
Moorwood and Oliva 1988; Oliva et al. 1994).
A strong 9.7 $\mu$m silicate absorption feature is seen
towards the nucleus (Moorwood and Glass 1984).
Br$\gamma$ imaging shows a compact central source
plus two extranuclear HII region complexes at a radius of 5$''$
(Moorwood and Oliva 1994b).
Circinus is unresolved in the IRAS 
CPC data with a resolution of $\sim$1$'$
(Ghosh et al. 1992; van Driel et al. 1993).
Our observations further confine the region of far-infrared emission
to a $\le$17$''$ $-$ $\le$28$''$ ($\le$330 $-$ $\le$540 pc) diameter area.

{\it 4.9. NGC 6946}

NGC 6946 is an Scd galaxy with 
a moderately active starburst nucleus (Keel 1984).
The molecular gas distribution in NGC 6946 is centrally peaked
(Young and Scoville 1982) and elongated in
a bar or spiral arm pattern (Ball et al. 1985;
Weliachew et al. 1988; Ishizuki et al. 1990; Regan and Vogel 1995),
while the atomic gas distribution has a central depression (Rogstad,
Shostak, and Rots 1973).

There have been several studies of the far-infrared morphology
of NGC 6946.
Low resolution 
IRAS maps
(Rice et al. 1988) give
a 60 $\mu$m radius of 13.3$'$ to $\mu$$_{60}$ = 25 mJy 
arcmin$^{-2}$.
With the KAO,
NGC 6946 was mapped at 
120 and 170 $\mu$m 
at 49$''$ resolution
(Smith, Harper, and Loewenstein 1984),
at 100 and
160 $\mu$m at 45$''$ resolution,
and at 200 $\mu$m at 56$''$ resolution
(Engargiola 1991). 
These maps show a central emission peak, an extended disk,
and some evidence for spiral structure in the disk emission.
Our higher resolution data set a limit 
of $\le$24$''$
$\times$ $\le$34$''$ ($\le$525 pc $\times$ $\le$710 pc)
FWHM on the size of the
central 100 $\mu$m source, consistent with the size of
the central CO (Ishizuki et al. 1990) and H$\alpha$ (Roy and Belley
1993) source.
We find that 20$\%$ of the total 100 $\mu$m flux density of NGC 6946 comes from
this central source.

{\it 4.10. NGC 7331}

NGC 7331 is an Sb galaxy 
with a prominent bulge (Boroson et al. 1981) and
a weak LINER-like nuclear optical emission line spectrum (Keel 1983a).
Narrow-band H$\alpha$+[N~II] imaging of
this galaxy (Keel 1983b; Pogge 1989) reveals a stellar-like nucleus
surrounded by diffuse ionized gas within the bulge and bright HII
regions in the spiral arms.
The diffuse bulge emission has line ratios similar to those
of the nucleus (Keel 1983b).
Radio continuum maps show a compact central source and faint diffuse
bulge emission, in addition to prominent spiral arms (Cowan,
Romanishin, and Branch 1994).
HI data show a central
depression in the atomic gas distribution (Bosma 1981).
Young and Scoville (1982) observed the major axis
of NGC 7331 in the CO (1 $-$ 0)
line at 45$''$ resolution and found a pronounced central hole;
their central position, however, 
was offset
19$''$ from the optical center. 
Subsequently, NGC 7331 was partially mapped in the CO (1 $-$ 0) and
(2 $-$ 1) lines at 23$''$ and 12$''$ resolution, respectively
(Braine et al. 1993; Braine 1995, private communication).
These data show a relatively constant CO (1 $-$ 0) surface brightness
over the central 1\arcmin.3 $\times$ 0\arcmin.5, with 
a possible 30$\%$ dip near the nucleus.  
In the higher resolution CO (2 $-$ 1) data, this dip is more pronounced,
with a
spiral arm/nucleus contrast of approximately 3.
Near-infrared data show redder colors at the
spiral arms than at the bulge, also suggesting a central 
dip in the distribution of dust and gas (Telesco, Gatley, and
Stewart 1982).

Our 100 $\mu$m major axis profile 
for NGC 7331 has a fairly flat-topped FIR distribution
with a FWHM of approximately 165$''$, without a pronounced central
depression or a bright compact nuclear source.
This far-infrared plateau covers not just the bulge
area, but also crosses the spiral arms at a radius of $\sim$50$''$.
We have also resolved the minor axis emission, with
a FWHM size of $\sim$42$''$.
This distribution is consistent with the lower resolution
100 $\mu$m IRAS CPC data of van Driel et al. (1993).

In Figure 5, we compare our 100 $\mu$m profile
with the corresponding CO (1 $-$ 0) 
data from Braine et al. (1993; 1995, private communication),
obtained with a similar beamsize.
Within the uncertainties, these 
distributions agree; 
our data do not rule out 
a 30$\%$ depression at 100 $\mu$m near the center of
NGC 7331.
In Figure 5, we have also plotted the major
axis H$\alpha$+[N II] profile for NGC 7331,
derived from the Pogge (1989) image.
In this figure, we plot an unsmoothed
cut across the image at the location of our
array, to identify the locations of the 
spiral arms and the bright compact
central source.
In addition, we plot the H$\alpha$+[N II]
profile after smoothing the image
to the same resolution
as our KAO data.
Since the emission
in the nuclear and 
bulge regions of this image is dominated
by [N~II] (Keel 1983a, 1983b)
and may not be due to OB star ionization,
to approximate the distribution of massive
young stars in NGC 7331
we have made a third cut across the
image, after removing this central emission and 
smoothing.
As this figure shows, within
our uncertainties the 100 $\mu$m
distribution is consistent with both
the original smoothed H$\alpha$+[N II] image
and the smoothed image with the bulge light
removed.  
Higher S/N and resolution are needed to unambiguously
distinguish between these two distributions.
NGC 7331 is highly inclined (75$^{\circ}$;
Bosma 1978), and 
the spiral arm HII regions along the minor axis
are partially contained within the central
beam.

{\it 4.11. NGC 7552}

The barred spiral galaxy NGC 7552 has an
`amorphous' nucleus
(S\'ersic and Pastoriza 1965),
with an emission line optical spectrum 
which has been
classified as HII region-like (Ward et al. 1980) or
as a weak LINER (Durret and Bergeron 1987, 1988).
Although diffuse 
ionized gas is 
observed to a radius of 1$'$, the bright central
region is relatively compact 
in H$\alpha$+[N~II],
with a FWHM size of $\sim$10$''$
(Durret and Bergeron 1987).
Radio and Br$\gamma$ 
maps of NGC 7552 reveal a nuclear starburst ring
of diameter 7$''$ $\times$ 9$''$ (Forbes et al. 1994;
Forbes, Kotilainen, and Moorwood 1994).
Our KAO data of this galaxy indicate that the far-infrared
emission is arising from an unresolved central source of size
$\le$22$''$ $\times$ $\le$24$''$, consistent with the 
size of this ring.

{\bf 5. Previously Published KAO Data}

To investigate the 
nature of far-infrared emission from galaxies, it
is important to make comparisons between different
galaxies as well as studies of individual galaxies.
We now have a large enough sample of galaxies
with high resolution KAO data available to make
this possible.

In this section we describe and
tabulate previously published KAO data on
11 additional galaxies.
These galaxies were observed by the University
of Texas KAO group using 
either one of the two detector systems described 
in Section 2, or a previous generation instrument.
Since these data
were 
obtained with either 
a scanning mode or with a multi-channel system, information
about the spatial extent of the central far-infrared
emission is available.
These additional galaxies are listed in Table 4, along with
their nuclear positions, distances, morphological
types, blue magnitudes, diameters, and
spectral types.
We also include in Table 4 the galaxies M 31 and M 81, which
are close enough that their bulge regions are resolved
by IRAS.

In Table 5, we summarize the KAO results for the galaxies
in Table 4, giving the wavelength
of the observations, the average position angle,
the source size, the observed flux density
(both peak and total, when available), and the literature reference.
We also include the total flux densities from IRAS,
and the Rickard and Harvey (1984) or other published small aperture
flux densities.

For all
24 galaxies in the combined
sample, in Table 6 we tabulate the 50 and 100 $\mu$m flux densities
that we use in the following discussion, along with the measured
far-infrared
source sizes in arcseconds 
and the physical area $\Omega$ in kpc$^2$.
Ideally, we would like to compare the far-infrared
emission from the same area within each galaxy,
the inner $\sim$1 kpc$^2$.
Since the galaxies are not all at the same distance, however,
this physical area varies from galaxy to galaxy.
Furthermore,
some of the galaxies 
are so distant that the
far-infrared size of the entire galaxy is used.

In Table 6, we also list the average dust temperature and
100 $\mu$m optical depth in the observed regions,
obtained from the 50 $\mu$m/100 $\mu$m flux ratios assuming
a $\lambda$$^{-1}$ emissivity law.
For galaxies without 50 $\mu$m measurements, 
we use the global IRAS 60 $\mu$m/100 $\mu$m ratio
to determine a dust temperature and therefore the
optical depth.  In these cases, 
if T$_d$(bulge) $>$ T$_d$(global)
then these estimates of $\tau$$_{100}$ are upper limits.
We also include the IRAS M 31 bulge flux densities for a 4$'$
aperture (Soifer
et al. 1986) in this table.
In addition,
we provide IRAS 60 and 100 $\mu$m measurements 
for the bulge of M 81 in a 105$''$ aperture obtained from
the IRAS HiRes images (see Table 6).

{\bf 6. Additional Data}

In Tables 7.1, 7.2, and 8, we provide
additional information about these
galaxies.
When possible,
we have obtained 
H$\alpha$+[N II]
fluxes 
for the regions observed by the KAO (Table 7.1).
The beamsizes for the 
H$\alpha$+[N II] measurements are also given in
Table 7.1.
For several of the galaxies in our sample, we
acquired copies of published H$\alpha$+[N II] images,
from which we have extracted aperture measurements.
For the other galaxies, we obtained fluxes
directly from the literature.
In these cases, the beamsizes are not always perfectly
matched with the KAO beam.
However, 
the optical apertures are generally
larger than those of the far-infrared, and
inspection of the published H$\alpha$+[N~II]
images shows that most of the
emission comes from the central far-infrared
source.
For most of these galaxies, the filters used for the
narrow-band optical observations were
wide enough that they contain both
H$\alpha$ and the nearby 
$\lambda$6584$\AA$ [N~II] line.
Exceptions are noted in Table 7.1.

We also give published Br$\gamma$ fluxes
in Table 7.1; these are only
sometimes useful, however, because
often the only
available Br$\gamma$ measurements are in apertures
significantly smaller than our KAO beam.
In Table 7.1, we also list published 
CO (1 $-$ 0) measurements obtained
in beam sizes similar to the KAO beam.
When necessary,
these have been corrected for the telescope aperture efficiency
to a main beam temperature scale (T$_{MB}$), suitable
for extended sources that fill the beam.
In the following comparison with the far-infrared
data, we use the first tabulated value of
these parameters
for
each galaxy.

For many of these galaxies, spectroscopic measurements
of the Balmer decrement F(H$\alpha$)/F(H$\beta$)
are also available.  These are given in Table 7.2,
along with their beamsizes, which are generally
much smaller than the KAO beamsize.
In addition, we also compile published values of
F(H$\alpha$)/F(Br$\gamma$) in Table 7.2.

We have also gathered BVH data for these
bulges at aperture sizes similar to the KAO beam,
using the NASA Extragalactic Database (NED; 1995)
and Longo and deVaucouleurs (1983, 1988).
From these data, we have estimated magnitudes
within our KAO beams assuming a linear log(aperture)-magnitude
relationship. 
If only one measurement at a given wavelength is available,
it is used in the subsequent analysis if the aperture
is within 30$\%$ of the KAO beam.
In Table 8, we tabulate the interpolated magnitudes 
we use in this study.
Finally,
in Table 9, we list some derived quantities for the observed
regions of these galaxies:
the far-infrared, H$\alpha$+[N II], and blue luminosities, 
along with the ratios L(FIR)/L(H$\alpha$+[N~II]), L(FIR)/L(B),
L(FIR)/$\Omega$, and L(B)/$\Omega$.

{\bf 6. Discussion}

{\it 6.1. Extinction}

To determine the origin of the far-infrared emission,
one must first calculate the extinction towards the
region in question.
Our data provides one such estimate, the 100 $\mu$m optical
depth (Table 6).  In Figure 6, we compare these values
with the integrated CO intensity I(CO), another measure
of the optical depth.
In this plot and in subsequent figures,
the galaxies with optical or near-infrared
signatures of strong star formation and/or Seyfert
activity within the KAO beam (see Tables 1 and 4, and
Section 4) are plotted as asterisks, while
the galaxies with more quiescent bulges are plotted as filled
triangles.  M 81, which has an extremely weak Seyfert nucleus,
is considered a quiescent galaxy.

Figure 6 shows that I(CO) and $\tau$$_{100}$ are
well-correlated, and
the galaxies with starburst or Seyfert activity
lie mostly in the upper righthand section of this plot, with higher
gas and dust column densities.
The line plotted in Figure 6
is the expected relationship if both
CO and $\tau$$_{100}$ trace extinction,
using 
A$_{\rm V}$/$\tau$$_{100}$ = 750 (Makinen et al. 1985) 
($\lambda$$^{-1}$ emissivity law) and
assuming the standard Galactic I(CO)/n(H$_2$) (Bloemen et al. 1986)
and N(H)/A$_{\rm V}$ (Bohlin, Savage, and Drake 1978) ratios.
The data points in Figure 6 generally lie above this
line, showing that the CO data imply optical
depths a factor of $\sim$4 higher than the far-infrared data
on average.
This discrepancy has been noted before in comparisons
of the 
global M(H$_2$) to M(dust) ratio (Young et al. 1989;
Devereux
and Young 1990) and in a comparison of single aperture far-infrared
data with published CO data (Rickard and Harvey 1984).
It may be due to a number of effects.
The A$_{\rm V}$/$\tau$$_{100}$ ratio is poorly known and
may vary significantly from location to location;
published values range from 280 $-$ 2600 (Harvey,
Thronson, and Gatley 1980; Whitcomb et al. 1981; Hildebrand
1983; Makinen et al. 1981; Casey 1991).
Also, extragalactic observations at 100 $\mu$m may miss a significant
amount of very cold dust (Eales,
Wynn-Williams, and Duncan 1989;
Kwan and Xie 1992).
In addition, the I(CO)/n(H$_2$)
ratio may be enhanced 
in starburst nuclei (Maloney
and Black 1988; Maloney 1990; Mauersberger et al.
1996a,b).
Scatter is also introduced by the differing
beamsizes, the lack of small
beam 50 $\mu$m data for many galaxies, and 
the possibility of clumpiness of the
interstellar medium within the KAO and CO beams.

Other methods of determining extinction 
are also
quite uncertain, particularly the broadband colors.
The B $-$ V color (Table 8.2) does not correlate with I(CO);
starburst and quiescent galaxies have similar colors.
B $-$ V does not trace
extinction well, because it is affected not only
by dust but also the stellar population.  
Also, because the dust is likely mixed with the
stars rather than in a foreground screen, the reddening
relative to the extinction is reduced (Witt,
Thronson, and Capuano 1992).
For a few individual galaxies,
detailed models 
of the optical
and near-infrared colors 
which take into account geometric effects
have recently been developed by a number of groups.
For NGC 4736, extinctions derived from
this method are consistent
with those obtained from the CO (1 $-$ 0) line (Block
et al. 1994).

Hydrogen emission line ratios may also underestimate
extinction because of mixing.
In Figure 7, we compare I(CO) 
with the A$_{\rm V}$ derived from the Balmer decrement.
For this calculation, we used the standard Galactic extinction
curve and an intrinsic F(H$\alpha$)/F(H$\beta$) = 2.85 (Osterbrock 1989),
except for the data for
the narrow line region of M 81, where an intrinsic
F(H$\alpha$)/F(H$\beta$)
= 3.1 was assumed (Filippenko and Sargent 1988).
This plot shows
little correlation, and the extinctions implied by
the Balmer decrement are generally smaller than those
given by I(CO).
The F(H$\alpha$)/F(Br$\gamma$) ratio is compared with
I(CO) in Figure 8.
There is somewhat better agreement,
but the hydrogen line ratio still gives lower extinction
values.
This may be due to a combination of underestimation
by F(H$\alpha$)/F(Br$\gamma$) due to mixing 
and overestimation by I(CO).

Other methods of determining extinction include
comparisons of these hydrogen lines with
the Pa$\beta$ and Br$\alpha$ 
lines, thermal radio continuum emission, 
submillimeter dust continuum, and mid-infrared absorption
feature strengths. Unfortunately, 
these data 
are available for only a few of our sample galaxies.
Therefore, in the following discussion, we will rely
on the optical depths determined from the CO intensities
and the 100 $\mu$m
optical depths, with the caveats noted above.
The true optical depth probably lies
between that implied by the CO data and 
that given by the far-infrared data.

{\it 6.2. Star Formation}

Galaxies which are actively forming stars
are known to have higher global far-infrared
luminosities and higher ratios of far-infrared
to blue luminosities
than less active galaxies.
The data tabulated in Table 9
show that
this is also true for 
the central regions of galaxies.
The
bulges of the quiescent galaxies, particularly M 31 and M 81,
stand
out as having low far-infrared luminosities 
relative to
their blue luminosities.
In Figure 9, we have
scaled these luminosities by the area $\Omega$ of the source,
to compare average surface brightnesses.
The starburst and Seyfert galaxies 
have higher
far-infrared surface brightnesses 
than the more quiescent galaxies.
Accounting for the fact that many of the starburst/Seyfert galaxies
are unresolved with the KAO makes this difference
more pronounced.
In this plot,
the blue surface brightnesses for the quiescent
and more active galaxies are similar, but these are underestimated
for the unresolved galaxies.

In Figure 10, far-infrared surface brightnesses are
compared to the CO (1~$-$~0) intensities.  A strong
correlation is seen, with higher
far-infrared surface brightnesses with increasing optical depth.
L(FIR)/L(B) is also correlated with
I(CO) (Figure 11).
Starburst/Seyfert galaxies have higher 
L(FIR)/L(B) ratios than the 
quiescent bulges.  

The H$\alpha$ luminosity, after
correction for extinction, is an approximate measure
of the rate of star formation, if ionization is
due to OB stars.
In Figures 12 and 13, 
we plot the ratio
L(FIR)/L(H$\alpha$) 
against our extinction measures I(CO) and $\tau$$_{100}$.
When necessary, we have
corrected for [N~II] emission
by assuming 
L(H$\alpha$) = 0.75L(H$\alpha$+[N~II]) for the 
galaxies with strong star formation within the KAO beam,
and 
L(H$\alpha$) = 0.25L(H$\alpha$+[N~II]) for the LINER galaxies
(see Jacoby et al. 1989;
Smith et al. 1991).
In contrast to Figure 11, where 
I(CO) is strongly correlated with
L(FIR)/L(B),
little 
correlation is found with L(FIR)/L(H$\alpha$).
The quiescent galaxies NGC 4736,
NGC 3627, and NGC 7331 have L(FIR)/L(H$\alpha$)
ratios higher than those of most of the starburst galaxies but
lower CO intensities and 100 $\mu$m optical depths.
In these figures, we plot curves corresponding
to the expected relationships for dust heating by
OB stars following a Salpeter IMF, 
using the assumptions given
above for extinction and an
intrinsic L(FIR)/L(H$\alpha$) ratio of
55
(Devereux and Young 1989).
For this calculation, we assume 
that the extincted region lies
halfway through the molecular
disk.

Although there is considerable uncertainty as to
the exact relationship between extinction, CO, and
$\tau$$_{100}$ (see Section 6.1),
these curves give an approximate dividing line
between galaxies in which dust heating is
dominated by OB stars and those in which the
older stellar population is important.
In Figure 12,
the quiescent galaxies tend to lie above the curve,
with infrared excesses higher than that expected by
dust heating by massive young stars.
The 
infrared excesses in the bulges of NGC 4736 and NGC 3627 
relative to their extinctions have
been noted before by
Smith et al. (1991, 1994), who conclude that 
dust heating is dominated by the non-ionizing stellar population.
This same conclusion was reached for 
the bulge of M 31 by
Soifer et al. (1986).
The central regions of NGC 7331 and M 81 may also
have major contributions by non-OB stars.

This analysis assumes that all of the observed H$\alpha$
in these galaxies is due to ionization by OB stars, 
however, this is probably not the case
for the bulges of the quiescent LINER-like galaxies.
The ionization mechanism responsible for the 
LINER-like optical line ratios in the centers
of galaxies is still unclear.
Shock excitation (Heckman 1980;
Bonatto et al. 1989), photoionization by
an active nucleus (Ferland and Netzer 1983), photoionization
of dense clouds by massive O stars (Filippenko and Terlevich
1992; Shields 1992), 
collisional excitation due to mass loss from giant stars
(Burbidge and Burbidge 1962;
Burbidge, Gould, and Pottash 1963), and
ionization by hot
horizontal-branch stars 
(Minkowski and Osterbrock 1959) have all been suggested
as possible ionization mechanisms.
If something other than OB stars is responsible for the
ionization, then the infrared excesses in these bulges
relative to the star formation rate is further enhanced.
In the bulges
of M 31 and NGC 4736, the optical line ratios
have been found to be LINER-like across the bulge, not
just in the nucleus (Ciardullo et al. 1988; Smith et al. 1994).

In Figure 12, the starburst
galaxies tend to lie to the right of the extinction curve,
therefore, there are sufficient OB stars in these
galaxies to power the far-infrared luminosity, if the CO
traces the extinction.
The fact that many of the starburst galaxies lie
far to the right in this plot may be due to an
overestimation of the H$\alpha$ extinction by the CO.
In contrast to Figure 12, in Figure 13 many of the starburst galaxies
lie to the left of the curve; this may be due to
an underestimation of A$_{\rm V}$ by $\tau$$_{100}$.

The peculiar galaxy NGC 1275 also lies above the
extinction curve, in the regime where dust heating by older stars 
is expected to dominate.
However, NGC 1275 may be a special case, being the dominant
galaxy in the X-ray luminous galaxy cluster Perseus.
The dust in this galaxy may be heated by hot intracluster
gas in a cooling flow rather
than star formation or the older
stellar population (Lester et al. 1995). 
Therefore
it
is unique in this sample, and we
do not discuss it further
in this paper.

These plots show that,
regardless of the precise relationship between I(CO),
$\tau$$_{100}$, and A$_{\rm V}$, the trends 
are clear; quiescent galaxies have higher far-infrared
excesses compared to starburst galaxies when extinction
corrections are taken into account.

In Figures 14 and 15, we compare L(FIR)/L(Br$\gamma$)
with I(CO) and $\tau$$_{100}$
for the galaxies for which beam-matched
Br$\gamma$ data 
are available.
All of these galaxies are known to have strong star formation
in the KAO beam,
and, in general, they fit the expected relationship
for OB star dust heating, with possible underestimation
of the optical depth by $\tau$$_{100}$ and overestimation
by I(CO).  
It is possible that an active galactic
nucleus contributes significantly to dust heating
in a few of these galaxies.
AGN dust heating is characterized by a high infrared
excess compared to
OB star heating (DePoy, Becklin, and Geballe 1987; 
Prestwich et al. 1994).
An infrared excess has been cited before for Arp 220,
based on extinctions derived from
F(Br$\alpha$)/F(Br$\gamma$), F(Br$\gamma$)/F(Pa$\gamma$),
and millimeter continuum data (DePoy et al.
1987; Scoville et al. 1991; Prestwich 
et al. 1994; Larkin et al. 1995; Armus et al. 1995).
The Seyfert nucleus in NGC 1068 may also contribute to
dust heating.
Further investigations of the extinction in these
galaxies are needed to confirm these results.

{\it 6.3. Dust Heating by the Older Stellar Population}

In Figure 11, we found that the far-infrared
to blue luminosity ratio increases with I(CO).
The blue light originates from a combination of both young and old
stars.  A better measure of the bolometric
luminosity of the older stellar population
is the near-infrared H band, which is also less affected
by extinction.
We compare the far-infrared to H-band luminosity
ratio with I(CO) in Figure 16.
As with L(FIR)/L(B), there is a correlation with
optical depth.  
Unlike the case with L(FIR)/L(H$\alpha$), there is
a clear distinction between starburst/Seyfert galaxies
and quiescent galaxies in L(FIR)/L(B) and L(FIR)/L(H),
with starburst/Seyferts having significantly higher values.

For comparison, we plot an extinction curve in
this figure, using
the same assumptions as above.
These curves are normalized to L(FIR)/L(H) $\sim$ 2
at A$_{\rm V}$ = 0, and pass near NGC 4736, NGC 3627,
and NGC 7331.  For a population of K5 giants,
this curve is consistent with
a situation where
$\sim$1/5th of the total
stellar luminosity 
is absorbed by dust
and re-emitted in the far-infrared.
These curves do not intersect M 31 and M 81,
implying that, if dust heating is
dominated by non-ionizing stars in all the quiescent
galaxies, then a higher proportion of the bolometric
luminosity of the older stars is absorbed by dust
and re-emitted in the far-infrared in NGC 4736, NGC 3267,
and NGC 7331 than in M 81 and M 81.
This difference may be caused by both a 
larger amount of interstellar dust available
for absorption and re-emission, which is clearly
the case,
and 
also a
difference in
stellar populations.
These three galaxies have younger stellar populations
and therefore more energetic radiation fields
than M 31 and M 81 (Pritchet 1977; Keel 1983c;
Walker, Lebofsky, and Rieke 1988).

In Figure 16,
the curve through NGC 4736 agrees with the data
for some of the starburst/Seyfert galaxies,
however, many lie
above the curve.
Therefore, 
the difference in L(FIR)/L(H) between starburst/Seyfert galaxies
and quiescent galaxies cannot simply be explained by extinction,
but instead must be due to differences in dust cross section
and stellar population.
The relative proportion of energy coming out of
the galaxy in the far-infrared relative to the optical/near-infrared
is higher in starburst/Seyfert galaxies.
Therefore, L(FIR)/L(old stars) is higher for starburst/Seyfert
galaxies than quiescent galaxies, 
while L(FIR)/L(young stars) is similar or lower (Section 6.2).

These results can be explained by an increasing
contribution to dust heating from young stars 
as one moves to higher extinctions.  
For the galaxies without optical/near-infrared signatures
of star formation, dust heating is dominated
by the underlying older stellar population.  For
more active galaxies, OB stars become important.
This can also be seen in Figure 17, where
L(H$\alpha$)/L(H) is plotted against I(CO).
L(H$\alpha$) measures
the 
number of observed OB stars (or,
as noted above,
for M 31, M 81, NGC 1275, and the LINERS, is an
upper limit), while L(H) traces the
older stellar population.
When the unusual galaxy NGC 1275 is excluded,
a segregation is seen between the 
quiescent galaxies, concentrated in
the lower left corner, and the starburst/Seyfert
galaxies, which have higher L(H$\alpha$)/L(H) ratios.
Thus, at larger column densities of gas and dust,
the proportion of young stars relative to
old stars is higher.

In these figures, the separation between the starbursts and
the more quiescent LINER bulges is at approximately log I(CO) = 1.5,
which corresponds to a molecular gas surface mass density
of 140 M$\sun$ pc$^{-2}$, using the Bloemen et al. (1986)
conversion.  This is 
the critical density at which star formation occurs 
in a gaseous disk 
with a rotational velocity of 150 km s$^{-1}$
at a radius of 500 pc (Kennicutt 1989).
Thus the lack of powerful star formation in the bulges
of NGC 4736, NGC 3627, NGC 7331, M 31, and M 81
may be due to an average gas density lower than
the critical density.

{\bf 7. Conclusions}

Using the KAO,
we have observed the central regions
of 11 galaxies 
in the far-infrared at high spatial resolution.
For 10 out of these 11 galaxies,
optical and near-infrared data indicate massive
star formation in the inner regions.
For these 10, the far-infrared sizes or size
limits are consistent with those of the
central star formation complexes.
The remaining galaxy, NGC 7331, does not have
observed star formation in its bulge.
It does not have a bright central far-infrared
source, but instead has a
very extended far-infrared distribution
with similar 100 $\mu$m surface brightnesses for
its bulge and inner spiral
arms.

We compare the small-beam far-infrared data for
these galaxies to optical, near-infrared,
and millimeter data, increasing the galaxy sample
to 24 by
including published KAO data for
11 additional galaxies and the
IRAS data for the bulges of M 31 and M 81.
We find
that L(FIR)/L(B) and L(FIR)/L(H) correlate strongly
with 
CO (1 $-$ 0) intensity and 100 $\mu$m
optical depth, while
L(FIR)/L(H$\alpha$) does not.
Galaxies with optical or near-infrared signatures of 
OB star formation 
in their central regions have higher values of I(CO) and $\tau$$_{100}$
than more quiescent galaxies,
as well as higher far-infrared surface brightnesses and 
L(FIR)/L(B) and L(FIR)/L(H) ratios.
The L(FIR)/L(H$\alpha$)
ratios for starburst galaxies
are consistent with OB stars dominating dust heating,
while 
more quiescent bulges have infrared excesses 
too high to be accounted for by young stars alone.
Whether or not a galaxy bulge has strong star formation,
and therefore far-infrared luminosities powered by
OB stars, may be decided by a threshold effect;
star forming galaxies have surface gas densities
above the Kennicutt critical density, while
quiescent galaxies have lower values.

\vskip 0.1in

We would like to thank 
J. Bergeron, 
J. Braine,
F. Casoli,
F. Durret, 
M. G\'erin,
R. Pogge, 
R. Rand, 
M. P. V\'eron,
and W. Waller
for providing copies of their data for our
use.
We would also like to thank 
Cecilia Colom\'e, James DiFrancesco, 
Marshall Joy,
Chris Koresko,
Dan Lester, and
Cheng-Yue Zhang 
for
their help in making these observations.
In addition, we acknowledge the help of Greg Doppmann
in calculating the asteroid fluxes.
We are very grateful to the KAO
staff for their support
during these flights, and to Dan Lester for helpful
advice throughout this project.
We also thank Sue Madden for helpful comments on
this manuscript.
This work was
supported by NASA grant NAG 2-67. 
B. J. S. also acknowledges the support of 
a National Research Council-JPL Research Associateship
during the final stages of this project.
This research has made use of the NASA/IPAC Extragalactic
Database (NED) which is operated by the Jet Propulsion
Laboratory, Caltech, under contract with the National
Aeronautics and Space Administration.

\vfill
\eject

{\bf REFERENCES}

\ref{Allen, R. J., Atherton, P. D., and Tilanus, R. P. J.
1986, Nature, 319, 296}

\ref{Aalto, S., Black, J. H., Booth, R. S., and Johansson, L. E. B.
1991a, A$\&$A, 247, 291}

\ref{Aalto, S., Black, J. H., Johansson, E. B., and Booth, R. S.
1991b, A$\&$A, 249, 323}

\ref{Antonucci, R. R. J., and Ulvestad, J. S. 1988, ApJ, 330, L97}

\ref{Armus, L., Heckman, T. M., and Miley, G. K. 1989, ApJ, 347, 727}

\ref{Armus, L., Neugebauer, G., Soifer, B. T., and Matthews, K. 
1995, AJ, 110, 2610}

\ref{Aumann, H. H., Fowler, J. W.,
and Melnyk, M. 1990, AJ, 99, 1674}

\ref{Barbon, R., Cappellaro, E., Ciatti, F., Turatto, M., and Kowal, C. T.
1984, A$\&$AS, 58, 735}

\ref{Beck, S. C., and Beckwith, S. V. 1984, MNRAS, 207, 671}

\ref{Beck, S. C., Turner, J. L., and Ho, P. T. P. 1986, ApJ, 309, 70}

\ref{Belley, J., and Roy, J.-R. 1992, ApJS, 78, 61}

\ref{Block, D. L., Witt, A. N., Grosbol, P., Stockton, A.,
and Moneti, A., 1994, A$\&$A, 288, 383}

\ref{Bloemen, J. B. G. L. et al. 1986, A$\&$A, 154, 25}

\ref{Bonatto, C., Bica, E., and Alloin, D. 1989, A$\&$A, 226, 23}

\ref{Bohlin, R. C., Corelli, R. H., Hill, J. K., Smith, A. M.,
and Stecher, T. P. 1983, ApJ, 274, L53}

\ref{Bohlin, R. C., Savage, B. D., and Drake, J. F. 1978, ApJ, 224, 132}

\ref{Booth, R. S. et al. 1989, A$\&$A, 216, 315}

\ref{Boroson, T. 1981, ApJS, 46, 177}

\ref{Bosma, A. 1978, Ph.D. Thesis, University of Groningen}

\ref{Bosma, A. 1981, AJ, 86, 1791}

\ref{Braine, J., Combes, F., Casoli, F., Dupraz, C., G\'erin, M.,
Klein, U., Wielebinski, R., and Brouillet, N. 1993, A$\&$AS, 97, 887} 

\ref{Brock, D., Joy, M., Lester, D. F., Harvey, P. M., and Ellis, 
H. B., Jr. 1988, ApJ, 329, 208} 

\ref{Burbidge, E. M., and Burbidge, G. R.  1962, ApJ, 135, 694}

\ref{Burbidge, G. R., Gould, R. J., and Pottasch, S. R. 1963, ApJ, 138, 945}

\ref{Buta, R. 1988, ApJS, 66, 233}

\ref{Canzian, B., Mundy, L. G., and Scoville, N. Z. 1988,
ApJ, 333, 157}

\ref{Campbell, M. F., Lester, D. F., Harvey, P. M.,
and Joy, M. 1989, ApJ, 345, 298}

\ref{Carlstrom, J. E., Jackson, J., Ho, P. T. P.,
and Turner, J. L. 1990, in The Interstellar Medium in
External Galaxies: Summaries of Contributed Papers,
(NASA CP-3084),
337}

\ref{Casey, S. C. 1991, ApJ, 371, 183}

\ref{Casoli, F., Combes, F., Augarde, R., Figon, P., and Martin, J. M.
1989, A$\&$A, 224, 31}

\ref{Casoli, F., Combes, F.,
Dupraz, C., G\'erin, M., Encrenaz, P. and Salez, M.
1988, A$\&$A, 192, L17}

\ref{Casoli, F., Dupraz, C.,
and Combes, F. 1992, A$\&$A, 264, 49}

\ref{Casoli, F., Dupraz, C., Combes, F., and Kaz\`es, I. A$\&$A,
1991, 251, 1}

\ref{Caulet, A. et al. 1992, ApJ, 388, 301}

\ref{Ciardullo, R., Rubin, V. C., Jacoby, G. H., Ford, H. C.,
and Ford, W. K., Jr. 1988, ApJ, 95, 438}

\ref{Claussen, M. J., and Shahai, R. 1992, AJ, 103, 1134}

\ref{Clements, E. D. 1981, MNRAS, 197, 829}

\ref{Condon, J. J., Condon, M. A., Gisler, G., and
Puschell, J. J. 1982, ApJ, 252, 102}

\ref{Corwin, H. G., Jr, 1991, Internal NED Report, `Positions
for Selected Extragalactic Objects in the NED Database'}

\ref{Cowan, J. J.,
Romanishin, W., and Branch, D. 1994, ApJ, 436, L139}

\ref{Coziol, R., Demers, S., Pena, M., and Barneoud, R. 1994, AJ, 108, 405}

\ref{Crane, P. C., Dickel, J. R., and Cowan, J. J. 1992, ApJ, 390, L9}

\ref{Dahari, O. 1985, ApJS, 57, 643}

\ref{Dahlem, M., Aalto, S., Klein, U., Booth,
R., Mebold, U., Wielebinski, R.,
and Lesch, H. 1990, A$\&$A, 240, 237}

\ref{Dame, T. M., Koper, E., Israel, F. P.,
and Thaddeus, P. 1993, ApJ, 418, 730}

\ref{de Vaucouleurs, G., and de Vaucouleurs, A.
1972, Mem. R. A. S., 77, 1}

\ref{de Vaucouleurs, G. 1975, in Stars and
Stellar Systems, Vol. 9, Galaxies and the Universe, 
ed. A. Sandage, M. Sandage, and J. Kristian (Chicago:
Univ. of Chicago Press), 309}

\ref{DePoy, D. L., Becklin, E. E., and Geballe, T. R. 1987,
ApJ, 316, L63}

\ref{DePoy, D. L. 1996, New Extragalactic Perspectives
in the New South Africa: Changing Perceptions of the
Morphology, Dust Content, and Dust-Gas Ratios in Galaxies,
ed. D. Block, in press}

\ref{Dettmar, R.-J. 1990, A$\&$A, 232, L15}

\ref{Devereux, N. A., and Young, J. S.
1990, ApJ, 359, 42}

\ref{Devereux, N. A., and Young, J. S.
1992, AJ, 103, 1536}

\ref{Devereux, N. A., and Young, J. S.
1993, AJ, 106, 948}

\ref{Devereux, N. A., and Scowen, P. A.
1994, AJ, 108, 1244}

\ref{Devereux, N. A., Price, R., Wells, L. A.,
and Duric, N. 1994, AJ, 108, 1667}

\ref{Devereux, N. A., Jacoby, G., and Ciardullo, R. 1995, AJ, 110, 1115}

\ref{Doyon, R., Joseph, R. D., and Wright, G. S. 1994, ApJ, 421, 101}

\ref{Dressel, L. L., and Condon, J. J. 1976, ApJS, 31, 187}

\ref{Deul, E. R. 1989, A$\&$A,
218, 78}

\ref{Durret, F., and Bergeron, J. 1987, A$\&$A, 173, 219}

\ref{Durret, F., and Bergeron, J. 1988, A$\&$AS, 75, 273}

\ref{Eales, S. A., Wynn-Williams, C. G.,
and Duncan, W. D. 1989, ApJ, 339, 859}

\ref{Eckart, A., Cameron, M., Rothermel, H., Wild, W., Zinnecker,
H., Rydbeck, G., Olberg, M., and Wiklind, T. 1990, ApJ,
363, 451}

\ref{Engargiola, G. 1991, ApJS, 76, 875}

\ref{Ferland, G. J., and Netzer, H. 1983, ApJ, 264, 105}

\ref{Filippenko, A. V., and Sargent, W. L. W.
1988, ApJ, 324, 134}

\ref{Filippenko, A. V., and Terlevich, R. 1992, 397, L79}

\ref{Fischer, J., Geballe, T. R., Smith, H. A., Simon, M.,
and Storey, J. W. V. 1987, ApJ, 320, 667}

\ref{Fischer, J., Simon, M., Benson, J., and Solomon, P. M.
1983, ApJ, 273, L27}

\ref{Forbes, D. A., Norris, R. P., Williger, G. M., and Smith, R. C.
1994, AJ, 107, 984}

\ref{Forbes, D. A., Kotilainen, J. K., and Moorwood, A. F. M.
1994, ApJ, 433, L13}

\ref{Ford, H. C., Crane, P. C.,
Jacoby, G. H., Lawrie, D. G.,
and van der Hulst, J. M. 1985, ApJ, 293, 132}

\ref{Freeman, K. C., Karlsson, B., Lynga, G., Burrell, J. F., van Woerden,
H., Goss, W. M., and Mebold, V. 1977, A$\&$A, 55, 445}

\ref{Gallais, P., Rouan, D., Lacombe, F., Tiph\`ene, D., and
Vauglin, I. 1991, A$\&$A, 243, 309}

\ref{Gallagher, J. S., Hunter, D. A.,
Gillett, F. C., and Rice, W. L. 1991, ApJ,
371, 142}

\ref{Garman, L. E., and Young, J. S. 1986, A$\&$A, 154, 8}

\ref{G\'erin, M., Nakai, N., and Combes, F. 1988, A$\&$A, 203, 44}

\ref{G\'erin, M., Casoli, F., and Combes, F. 1991, A$\&$A, 251, 32}

\ref{Ghosh, S. K., Bisht, R. S., Iyengar, K. V. K., Rengarajan, T. N.,
Tandon, S. N., and Verma, R. P. 1992, ApJ, 391, 111}

\ref{Goad, J., and Gallagher, J. S., Jr. 1985,
ApJ, 297, 98}

\ref{Graham, J. R. 1979, ApJ, 232, 60}

\ref{Graham, J. R., Wright, G. S., Meikle, W. P. S., Joseph, R. D.,
and Bode, M. F. 1984, Nature, 310, 213}

\ref{Graham, J. R., Wright, G. S., Joseph, R. D., Frogel, J. A.,
Phillips, M. M., and Meikle, W. P. S. 1987, in Star Formation in
Galaxies, ed. C. J. Lonsdale Persson (NASA CP-2466), 517}

\ref{Hall, D. N. B., Kleinmann, S. G., Scoville, N. Z., and
Ridgway, S. T. 1981, ApJ, 248, 898}

\ref{Handa, T., Nakai, N., Sofue, Y., Hayashi, M., and Fujimoto, M. 1990,
PASJ, 42, 1}

\ref{Harper, D. A., and Low, F. J. 1973, ApJ, 182, L89}

\ref{Harvey, P. M., and Wilking, B. A. 1982, PASP, 94, 285}

\ref{Harvey, P. M., Hoffmann, W. F., and Campbell, M. F.
1978, 70, 165}

\ref{Harvey, P. M., Lester, D. F., Brock, D., and Joy, M.
1991, ApJ, 368, 558}

\ref{Harvey, P. M., Thronson, H. A., and Gatley, I. 1980,
ApJ, 235, 894}

\ref{Heckman, T. M., Balick, B., and
Crane, P. C. 1980, A$\&$AS, 40, 295} 

\ref{Heckman, T. M. 1980, A$\&$A, 87, 152}

\ref{Helfer, T. T., and Blitz, L. 1993, ApJ, 419, 86}

\ref{Helou, G. 1986, ApJ, 311, L33}

\ref{Hildebrand, R. H. 1983, QJRAS, 24, 267}

\ref{Ho, P. T. P.,
Beck, S. C., and Turner, J. L.
1990, ApJ, 349, 57}

\ref{Ho, L. C., Filippenko, A. V., and Sargent, W. L. W. 1995,
ApJS, 98, 477}

\ref{Hummel, E., van der Hulst, J. M., and Keel, W. C. 1987, A$\&$A,
172, 32}

\ref{Humphreys, R. M., and Aaronson, M. 1987, AJ, 94, 1156}

\ref{Hutchings, J. B., Neff, S. G., Stanford, S. A.,
Lo, E., and Unger, S. W. 1990, AJ, 100, 60}

\ref{IRAS Catalogs and Atlases: Explanatory Supplement 1988,
ed. C. A. Beichman, G. Neugebauer, H. J. Habing,
P. E. Clegg, and T. J. Chester (Washington DC: GPO)}

\ref{IRAS Minor Planet Survey, 1992, edited by Tedesco,
E. F. (Phillips Laboratory Technical Report No. PL-TR-92-2049,
Hanscom Air Force Base, MA)}

\ref{Israel, F. P., van Dishoeck, E. F., Baas, F.,
Koornneef, J., Black, J. H., and
de Graauw, T. 1990, A$\&$A, 227, 342}

\ref{Jacoby, G. H., Ford, H., and Ciardullo, R. 1985, ApJ, 290, 136}

\ref{Joseph, R. D. and Wright, G. S. 1985, MNRAS, 214, 87}

\ref{Joy, M., Lester, D. F., and Harvey, P. M. 1987, ApJ,
319, 314}

\ref{Joy, M., Lester, D. F., Harvey, P. M., and Ellis,
H. B. 1988, ApJ, 326, 662}

\ref{Joy, M., Lester, D. F., Harvey, P. M., and Frueh, M.
1986, ApJ, 307, 110}

\ref{Joy, M., Lester, D. F., Harvey, P. M., Telesco, C. M.,
and Decher, R. 1989, ApJ, 339, 100}

\ref{Kawara, K., Nishida, M., and Gregory, B. 1987, ApJ, 321, L35}

\ref{Keel, W. C. 1983a, ApJS, 52, 229}

\ref{Keel, W. C. 1983b, ApJ, 268, 632}

\ref{Keel, W. C. 1983c, ApJ, 269, 466}

\ref{Keel, W. C. 1984, ApJ, 282, 75}

\ref{Keel, W. C. 1989, AJ, 98, 195}

\ref{Kenney, J. D. P., Scoville, N. Z., and Wilson, C. D. 
1991, ApJ, 366, 432}

\ref{Kennicutt, R. C., Jr. 1988, ApJ, 334, 144}

\ref{Kennicutt, R. C., Jr. 1989, ApJ, 344, 685}

\ref{Kennicutt, R. C., and Kent, S. M. 1983, AJ, 88, 1094}

\ref{Kent, S. M. 1985, ApJS, 59, 115}

\ref{Kinney, A. L., Bohlin, R. C., Calzetti, D., Panagia,
N., and Wyse, R. F. G. 1993, ApJS, 86, 5}

\ref{Krabbe,
A., Stenberg, A.,
and Genzel, R.
1994, ApJ, 425, 72}

\ref{Kr\"ugel, E., Chini, R., Klein, U., Lemke, R., Wielebinski, R.,
and Zylka, R. 1990, A$\&$A, 240, 232}

\ref{Kuhr, H., Witzel, A., Pauliny-Toth, I. I. K., and Nauber, U.
1981, A$\&$AS, 45, 367}

\ref{Kwan, J., and Xie, S. 1992, ApJ, 398, 105}

\ref{Larkin, J. E., Graham, J. R.,
Matthews, K., Soifer, B. T.,
Beckwith, S., Herbst, T. M.,
and Quillen, A. C.
1994, ApJ, 420, 159}

\ref{Larkin, J. E., Armus, L., Knop, R. A., Matthews, K.,
and Soifer, B. T. 1995, ApJ, 452, 599}

\ref{Lauberts, A. 1982, `The ESO/Uppsala Survey of the ESO(B) Atlas',
Munich, European Southern Observatory}

\ref{Lester, D. F. 1995, in Proceedings of the Airborne
Astronomy Symposium on the Galactic Ecosystem: From
Gas to Stars to Dust, ed. M. R. Haas, J. A. Davidson,
and E. F. Erickson, A.S.P. Conference Series,
Volume 73 (San Francisco: BookCrafters, Inc),
159}

\ref{Lester, D. F., Harvey, P. M., and Joy, M. 1986a, ApJ,
302, 280}

\ref{Lester, D. F., Harvey, P. M., and Joy, M. 1986b, ApJ,
304, 623}

\ref{Lester, D. F., Joy, M., Harvey, P. M., Ellis, H. B.,
and Parmar, P. S. 1987, ApJ, 321, 755}

\ref{Lester, D. F., Zink, E. C., Doppmann, G. W.,
Gaffney, N. I., Harvey, P. M., Smith, B. J.,
and Malkan, M. 1995, ApJ, 439, 185}

\ref{Lord, S. D., and Young, J. S. 1990, ApJ, 356, 135}

\ref{Lord, S. D. 1987, Ph.D. Dissertation, University of Massachusetts}

\ref{Nakagawa, T., Nagata, T., Geballe, T. R., Okuda, H.,
Shibai, H., and Matsuhara, H. 1989, ApJ, 340, 729}

\ref{Neff, S. G., and Hutchings, J. B. 1992, AJ, 103, 1746}

\ref{Makinen, P., Harvey, P. M., Wilking, B. A., and Evans, N. J.
1985, ApJ, 299, 341}

\ref{Maloney, P., and Black, J. H. 1988, ApJ, 325, 389}

\ref{Maloney, P. 1990, The Interstellar Medium in Galaxies,
ed. H. Thronson and M. Shull (Boston: Kluwer), 493}

\ref{Marsh, K. A., and Helou, G. 1995, ApJ, 445, 599}

\ref{Marston, A. P. 1989, AJ, 98, 1572}

\ref{Mauersberger, R., Henkel, C., Wielebinski, R.,
Wiklin, T., and Reuter, H.-P. 1996a, A$\&$A, in press}

\ref{Mauersberger, R., Henkel, C., Whiteoak, J. B.,
Chin, Y.-N., and Tieftrunk, A. R. 1996b, A$\&$A, in press}

\ref{Meaburn, J., Terrett, D. L., Teokas, A., and Walsh, J. R.
1981, MNRAS, 195, 39}

\ref{Minkowski, R., and Osterbrock, D. 1959, ApJ, 129, 583}

\ref{Mirabel, I. F.,
Booth, R. S., Garay, G., 
Johansson, L. E. B.,
and Sanders, D. B.
1990, A$\&$A, 236, 327}

\ref{Moorwood, A. F. M., and Glass, I. S. 1984, A$\&$A, 135, 281}

\ref{Moorwood, A. F. M., and Oliva, E. 1988, A$\&$A,
203, 279}

\ref{Moorwood, A. F. M., and Oliva, E. 1994a, ApJ, 429, 602}

\ref{Moorwood, A. F. M., and Oliva, E. 1994b, Infrared Phys. Tech.,
35, 349}

\ref{Morganti, R., Robinson, A., Fosbury, R. A. E.,
di Serego Alighieri, S., Tadhunter, C. N., and Malin, D. F.
1991, MNRAS, 249, 91}

\ref{Morgan, W. W. 1958, PASP, 70, 364}

\ref{Nilson, P. 1973, Uppsala General Catalogue of Galaxies,
Uppsala, Sweden: Royal Society of Sciences of Uppsala)}

\ref{Norris, R. P. 1985, MNRAS, 216, 701}

\ref{Oliva, E., and Moorwood, A. F. M.
1990, ApJ, 348, L5}

\ref{Oliva, E., Salvati, M., Moorwood, A. F. M.,
and Marconi, A. 1994, A$\&$A, 288, 457}

\ref{Osmer, P. S., Smith, M. G., and Weedman, D. W. 1974, ApJ, 192, 279}

\ref{Osterbrock, D. E. 1989, Astrophysics of Gaseous Nebulae and Active
Galactic Nuclei (Mill Valley, CA: University Science Books}

\ref{Patnaik, A. R. et al. 1992, MNRAS, 254, 655}

\ref{Peimbert, M., and Spinrad, H. 1970, ApJ, 160, 429}

\ref{Peimbert, M. and Torres-Peimbert, S. 1981, AJ, 245, 845}

\ref{Pence, W. D. 1980, ApJ, 239, 54}

\ref{Persson, C. J. L., and Helou, G. 1987, ApJ, 314, 513}

\ref{Phillips, A. C. 1993, AJ, 105, 486}

\ref{Pogge, R. W. 1989, ApJS, 71, 433}

\ref{Prestwich, A. H., Joseph, R. D., and Wright, G. S. 1994,
ApJ, 422, 73}

\ref{Pritchet, C. 1977, ApJS, 35, 397}

\ref{Puxley, P. J., Hawarden, T. G., and Mountain, C. M.
1988, MNRAS, 234, 29P}

\ref{Rand, R. J.,
Kulkarni, S. R., and Rice, W.
1992, ApJ, 390, 66}

\ref{Reynolds, J. E. et al. 1994, AJ, 108, 725}

\ref{Rice, W., Boulanger, F., Viallefond, F.,
Soifer, B. T., and Freedman, W. L. 1990, ApJ, 358, 418}

\ref{Rice, W., Lonsdale, C. J., Soifer, B. T., Neugebauer, G., Koplan, E. L.,
Lloyd, L. A., de Jong, T., and Habing, H. J. 1988, ApJS, 68, 91}

\ref{Rice, W. 1993, AJ, 105, 67}

\ref{Rickard, L. J., and Harvey, P. M. 1984, AJ, 89, 1520}

\ref{Rieke, G. H., Cutri, R. M., Black, J. H.,
Kailey, W. F.,
McAlary, C. W., Lebofsky, M, J., and Elston,
R. 1985, ApJ, 290, 116}

\ref{Rieke, G. H., Lebofsky, M. J.,
Thompson, R. I., Low, F. J., and
Tokunaga, A. 1980, ApJ, 238, 24}

\ref{Rieke, G. H., Lebofsky, M. J.,
and Walker, C. E. 1988, ApJ, 325, 679}

\ref{Rose, J. A., and Searle, L. 1982, ApJ, 253, 556}

\ref{Rotaciuc, V., Krabbe, A., Cameron, M., Drapatz,
S., Genzel, R., Sternberg, A., and Storey, J. W. V. 1991,
ApJ, 370, L23}

\ref{Rotaciuc, V., Krabbe, A., Cameron, M., Drapatz,
S., Genzel, R., Sternberg, A., and Storey, J. W. V. 1991,
ApJ, 370, L23}

\ref{Rowan-Robinson, M. 1986, MNRAS, 219, 737}

\ref{Roy, J.-R., and Belley, J. 1993, ApJ, 406, 60}

\ref{Rupen, M. P. 1991, AJ, 102, 48}

\ref{Sage, L. J., and Isbell, D. W. A$\&$A, 247, 320}

\ref{Sage, L. J., and Westpfahl, D. J. A$\&$A,
242, 371}

\ref{Saikia, D. J., Unger, S. W., Pendlar, A., Yates, G. J.,
Axon, D. J., Wolstencroft, R. D., Taylor, K., and Gyldenkerne, K. 
1990, MNRAS, 245, 397}

\ref{Saikia, D. J., Pedlar, A., Unger, S. W., and Axon,
D. J. 1994, MNRAS, 270, 46}

\ref{Sandage, A., and Brucato, 1979, AJ, 84, 472}

\ref{Sandage, A., and Tammann, G. A. 1981, A Revised
Shapley-Ames Catalog of Bright Galaxies (Washington DC:
Carnegie Institution of Washington)}

\ref{Sanders, D. B., and Mirabel, I. F. 1985, ApJ, 298, L31}

\ref{Sargent, A. I., Sanders, D. B., and Phillips, T. G.
1989, ApJ, 346, L9}

\ref{Schwering, P. B. W., and Israel, F. P. 1989,
A$\&$AS, 79, 79}

\ref{Schwering, P. B. W. 1989, A$\&$AS, 79, 105}

\ref{Scoville, N. Z., Sargent, A. I., Sanders, D. B., and Soifer, B. T.
1991, ApJ, 366, L5}

\ref{Scoville, N. Z., Soifer, B. T., Neugebauer, G., Young,
J. S., Matthews, K., and Yerka, J. 1985, ApJ, 289, 129}

\ref{Scoville, N. Z., Thakkar, D., Carlstrom, J. E., and Sargent, A. I.
1993, ApJ, 404, L59}

\ref{Scoville, N. Z., and Young, J. S. 1983, ApJ, 265, 148}

\ref{Seyfert, C. K. 1943, ApJ, 97, 28}

\ref{Shield, J. C. 1992, ApJ, 399, L27}

\ref{Simien, F., and de Vaucouleurs, G. 1986, ApJ, 302, 564}

\ref{S\'ersic, J. L., and Pastoriza, M. 1965, PASP, 77, 287}

\ref{Smith, B. J., Lester, D. F., Harvey, P. M., and Pogge, R. W. 1991, 
ApJ, 373, 66}

\ref{Smith, B. J., Harvey, P. M., Colom\'e, C., Zhang, C. Y.,
Di Francesco, J., and Pogge, R. W. 1994, ApJ, 425, 91}

\ref{Smith, B. J., Harvey, P. M., and Lester, D. F. 1995, ApJ,
442, 610}

\ref{Smith, J. 1982, ApJ, 261, 463}

\ref{Smith, J., Harper, D. A., and Loewenstein, R. F.
1984, in Proc. Airborne Astronomy Symposium, ed. H. A. Thronson, Jr. and
E. F. Erickson (NASA CP-2353), 277}

\ref{Stier, M. T., Traub, W. A., Fazio, G. G., Wright, E. L.,
and Low, F. J. 1978, ApJ, 226, 347}

\ref{Storchi-Bergmann, T., Kinney, A. L., and Challis, P. 1995, 
ApJS, 98, 103}

\ref{Surace, J. A., Mazzarella,
J., Soifer, B. T., and Wehrle,
A. E. 1993, AJ, 105, 864}

\ref{Tacconi, L. J. and Young, J. S.
1989, ApJS, 71, 455}

\ref{Talbot, R. J., Jr., Jensen, E. B., and Dufour, R. J. 1979,
ApJ, 229, 91}

\ref{Telesco, C. M., and Gatley, I. 1981, ApJ, 247, L11}

\ref{Telesco, C. M., and Harper, D. A. 1980, ApJ, 235, 392}

\ref{Telesco, C. M., Becklin, E. E., Wynn-Williams, C. G.,
and Harper, D. A. 1984, ApJ, 282, 427}

\ref{Telesco, C. M., Dressel, L. L., and Wolstencroft, R. D.
1993, ApJ, 414, 120}

\ref{Telesco, C. M., Gatley, I., and Stewart, J. M. 1982, ApJ, 263, L13}

\ref{Telesco, C. M., Harper, D. A., and Loewenstein, R. F. 1976,
ApJ, 203, L53}

\ref{Thompson, R. I., Lebofsky, M. J., and Rieke, G. H.
1978, ApJ, 222, L49}

\ref{Turner, J. L., and Ho, P. T. P. 1994, ApJ, 421, 122}

\ref{Turner, J. L., Ho, P. T. P., and Beck, S. C.
1987, ApJ, 313, 644}

\ref{Walterbos, R. A. M., and Schwering, P. B. W. 1987,
A$\&$A, 180, 27}

\ref{Wainscoat, R. J., de Jong, T., and Wesselius, P. R. 1987, A$\&$A,
181, 225}

\ref{Walker, C. E., Lebofsky, M. J., and Rieke, G. H.
1988, ApJ, 325, 687}

\ref{Wall, W. F., Jaffe, D. T., Israel, F. P., and Bash, F. N.
1991, ApJ, 380, 384}

\ref{Waller, W. H., Kleinmann, S. G., and Ricker, G. R. 1988,
AJ, 95, 1057}

\ref{Wamsteker, W., Prieto, A., Vitores, A., Schuster, H. E.,
Danks, A. C., Gonzales, R., and Rodrigues, 
1985, A$\&$AS, 62, 255}

\ref{Ward, M., Penston, M. V., Blades, J. C., and Turtle, A. J.
1980, MNRAS, 193, 563}

\ref{Weliachew, L., Casoli, F., and Combes, F. 1988, A$\&$A,
199, 29}

\ref{Whitcomb, S. E., Gatley, I., Hildebrand, R. H., Keene, J.,
Sellgren, K., and Werner, M. W. 1981, ApJ, 248, 416}

\ref{Willner, S. P., Soifer, B. T., Russell, R. W., Joyce, R. R.,
and Gillett, F. C. 1977, ApJ, 217, L121}

\ref{Witt, A. N., Thronson, H. A., and Capuano, J. M., Jr.
1992, ApJ, 393, 611}

\ref{Wynn-Williams, C. G., Becklin, E. E.,
Matthews, K., and Neugebauer, G. 1977, MNRAS, 189, 163}

\ref{Wynn-Williams, C. G., Becklin, E. E., Mathews, K., and Neugebauer,
G. 1979, MNRAS, 189, 161}

\ref{van Driel, W. et al. 1995, AJ, 109, 942}

\ref{van Driel, W., de Graaum, Th., de Jong, T., and Wesselius, P. R.
1993, A$\&$AS, 101, 207}

\ref{Van Dyk, S. D., Hyman, S. D., Sramek, R. A., and Weiler, K. W. 1994,
IAU Circular 6045, 1}

\ref{V\'eron-Cetty, M.-P., and V\'eron, P. 1985, A$\&$A, 145, 425}

\ref{V\'eron-Cetty, M.-P., and V\'eron, P. 1986, A$\&$AS, 66, 335}

\ref{Vigotti, M., Grueff, G., Perley, R., Clark, B. G.,
and Bridle, A. H. 1989, AJ, 98, 419}

\ref{Young, J. S. et al. 1995, ApJS, 98, 219}

\ref{Young, J. S., Kleinmann, S. G., and Allen, L. E.
1988, ApJ, 334, L63}

\ref{Young, J. S., Schloerb, F. P., Kenney, J., and Lord, S. D.
1986, ApJ, 304, 443}

\ref{Young, J. S., and Scoville, N. 1982, ApJ, 260, L41}

\ref{Young, J. S., Tacconi, L. J., and Scoville, N. Z. 1983, ApJ,
269, 136}

\ref{Young, J. S., Xie, S., Kenney, J. D. P., and Rice, W. L. 1989,
ApJS, 70, 699}

\vfill
\eject

{\bf CAPTIONS}

Figure 1.  The co-added scan data.  
For the NGC 3256 and Uranus 5/23/90 data,
which were acquired with the 20 channel system,
only the data from the central 8 channels in
the first bank of detectors are displayed.
The orientation of the array on the sky and the
scan direction are indicated
by the coordinates on the plots and by the rotation
angle in Table 2.
The detectors are separated by 13\arcsec.8 at 100 $\mu$m
and 7\arcsec.0 at 50 $\mu$m.

Figure 2.  The galaxy scan data for the peak channel,
plotted with the data for the point source object from
the same flight.
The curves with the larger error bars are the galaxy profiles.

Figure 3.  Far-infrared profiles along the long axis of the array,
obtained from averaging the central 22$''$ (100 $\mu$m)
or 10$''$ (50 $\mu$m) of the summed scans for each detector.
The data for the point source object from the
same flight (open squares) is displayed with the galaxy
data (filled circles).

Figure 4.  The nod data.  The top panel displays the data
from the first bank of detectors, while the lower panel
shows the data from the second bank.
Both the galaxy data and the corresponding point source
profile are shown.
In Figure 4d, we display the same M 83 and $\eta$ Carinae data 
that are shown in Figure 4c, but 
with the end channels set to zero.
During the NGC 891 observations, the array was shifted along its minor axis,
giving two overlapping sets of data.
For NGC 7331, three overlapping sets of data were obtained.

Figure 5.  The major axis 100 $\mu$m profile for NGC 7331,
along with the 23$''$ resolution CO (1 $-$ 0) observations
of Braine et al. (1993; 1995, private communication).
The (0,0) point of the Braine et al. CO data is offset from the
optical and radio continuum of NGC 7331 (Table 1); from
their partial map, we have selected
the positions closest to our observed locations for comparison (within
5$''$).  This plot also contains 3 profiles derived from
the Pogge (1989) H$\alpha$+[N II] image: 1) a cut 
across the unsmoothed array along
our array position, 2) a cut along our array position after the
image has been smoothed to our beamsize, and 3) a cut along our
array location after the central bulge and nuclear emission
has been removed and the image smoothed.

Figure 6. Log I(CO) vs. Log $\tau$$_{100}$.
The data for the galaxies with strong
star formation and/or Seyfert activity within the KAO beam
are plotted as asterisks, while the more quiescent galaxies are
plotted as filled triangles.
The plotted curves show expected relationships
using the standard I(CO)/n(H$_2$) ratio (Bloemen
et al. (1986) and the Makinen et al. (1985)
A$_{\rm V}$/$\tau$$_{100}$ relationship (see text).
Note that, for the galaxies without small beam 50 $\mu$m
measurements and T$_d$(bulge) $>$ T$_d$(global),
the plotted values of $\tau$$_{100}$ are upper 
limits (see Table 6).

Figure 7. Log I(CO) vs. the log A$_{\rm V}$ derived from
the Balmer decrement.
Symbols are as in Figure 6.

Figure 8. Log I(CO) vs. the log A$_{\rm V}$ derived from
F(H$\alpha$)/F(Br$\gamma$).
Symbols are as in Figure 6.

Figure 9. Log L(B)/$\Omega$ vs. Log L(FIR)/$\Omega$.
Symbols are as in Figure 6.

Figure 10. Log L(FIR)/$\Omega$ vs. Log I(CO).
Symbols are as in Figure 6.

Figure 11. Log L(FIR)/L(B) vs. Log I(CO).
Symbols are as in Figure 6.

Figure 12. Log L(FIR)/L(H$\alpha$) vs. Log I(CO).
Symbols are as in Figure 6.
The plotted curve is the line expected for dust
heating by OB stars, assuming a Salpeter IMF and
the standard Galactic I(CO)/n(H$_2$) ratio (Bloemen et al. 1986).

Figure 13. Log L(FIR)/L(H$\alpha$) vs. $\tau$$_{100}$.
Symbols are as in Figure 6.
The plotted curve is the line expected for dust
heating by OB stars, assuming a Salpeter IMF and
A$_{\rm V}$/$\tau$$_{100}$ (Makinen et al. 1985).

Figure 14. Log L(FIR)/L(Br$\gamma$) vs. I(CO).
Symbols are as in Figure 6.
The plotted curve is the line expected for dust heating
by OB stars, as in Figure 12.

Figure 15. Log L(FIR)/L(Br$\gamma$) vs. $\tau$$_{100}$.
Symbols are as in Figure 6.
The plotted curve is the line expected for dust heating
by OB stars, as in Figure 13.

Figure 16. Log L(FIR)/L(H) vs. I(CO).
Symbols are as in Figure 6.
The solid curve is the relationship expected by extinction
using the assumptions given in the text.
It is 
normalized to L(FIR)/L(H) $\sim$ 2 at A$_{\rm V}$ = 0,
which
corresponds to a situation where $\sim$1/5th
of the total bolometric luminosity of the stars
is absorbed by dust
and re-emitted in the far-infrared, assuming a K5 giant
stellar population.

Figure 17. Log L(H$\alpha$)/L(H) vs. I(CO).
Symbols are as in Figure 6.

\end

To obtain the most spatial information possible, we have
applied the iterative 1-dimensional maximum entropy program described
in Lester, Harvey, and Joy (1986a) to the data for the brightest
channel for each galaxy.  
These results are consistent with the deconvolved data to a $\chi$$^2$
= N level.
For the deconvolutions, a 20$''$ portion
at each end of the scans was omitted.
The FWHM of the deconvolved profiles are 
also given in Table 3, and are consistent within 5$''$ with
that derived from a simple gaussian deconvolution.
The approximate spatial resolution reached is 
8$''$ at
50 $\mu$m and 12$''$ at 100 $\mu$m.

[ NGC 1097  ... , and a possible
(2$\sigma$) feature extending to the southeast.
[  check this..delete this part?  ]

NGC 1097: 
To compare the spatial distribution of the far-infrared with the stellar
and ionized gas distribution, W. Keel has kindly provided us with
a copy of their H$\alpha$ image of NGC 1097 (Hummel, van der
Hulst, and Keel 1987).
We transformed this data into a form suitable for comparison with
the far-infrared data, by constructing a ``synthetic scan."  This was
created by rotating the image, taking a 3.0$'$ cut, and weighting the data in
the cross-scan direction by a Gaussian of FWHM = 25$''$, 
to approximate the response of the FIR detector.
This scan was then smoothed to 12$''$ resolution.  
The result is shown in Figure 4a, where it is compared with
the 100 $\mu$m deconvolved data.  
This figure shows that the 
the FIR, with the exception of the 2$\sigma$ wing, traces the H$\alpha$
distribution well.
The ring feature seen in the H$\alpha$ image is smoothed out at 12$''$
resolution, and is no longer seen.  
The close correlation between the FIR and the unobscured H$\alpha$ suggests
that the FIR in NGC 1097 is mainly powered by star formation in the ring.
\vfill\eject

A comparison of the NGC 1808
data with the Ceres data shows a possible extension beyond the point source,
at the 2$\sigma$ level in the wings. 

`Synthetic scans' for the V/'eron-Cetty and Veron (1985) B, r, and z 
($\lambda$$_{eff}$ $\sim$ 9000$\AA$)
images were made for comparison with the FIR, from
the copies of the images M. Veron kindly gave us.
These are shown in Figure 4b.
These show that the far-infrared traces the stellar light well,
within the resolution limit of the KAO data.

Circinus:
Using F($\lambda$) = B$_{\lambda}$$(T_d)$[1 - e$^{-{\tau}_{\lambda}}$]$\Omega$,
where F$_{\lambda}$ is the flux density at 50 or 100 $\mu$m, 
B$_{\lambda}$$(T_d)$ is a blackbody, and $\Omega$ is
the solid angle of the source, we solved for T$_d$ and 
$\tau$$_{100 {\mu}m}$, assuming a $\lambda$$^{-1}$ emissivity law. 
We find a dust temperature of 38K and an 100 $\mu$m optical depth
of $\ge$0.12 $\pm$ 0.08.  Using A$_v$ = 500-1000 $\tau$$_{100 {\mu}m}$
(Makinen et al. 1985), this implies that the visual extinction
is A$_v$ $\ge$ 18.  For dust particles in radiative equilibrium surrounding
a point source, r$^2$ $\le$ 3969L(L$\sun$)T$^{-5}$ pc$^2$, where
r is the radius of the far-infrared emitting region,
L is the luminosity of the central source, and T is the dust temperature
in degrees Kelvin (Joy et al. 1988).  
This relationship gives an upper limit to
the radius of 640 pc, greater than
our observed radius.  
Thus a central point source of this luminosity is capable of heating 
dust to our observed radius.
These calculations show that the source which
is powering the far-infrared is highly obscured, and the far-infrared
source size is not inconsistent with the heating being due
to a central point
source.

NGC 7552:
We constructed synthetic H$\alpha$ and red continuum
scans from an H$\alpha$ image and a red continuum image
kindly provided by F. Durret and J. Bergeron (from Durret
and Bergeron 1987).  This is compared with the deconvolved 100 $\mu$m
scan in Figure 4f.  This figure shows that the far-infrared 
profile is somewhat
wider than the H$\alpha$ profile, more in agreement with the red profile.
This suggests that some contributions to the far-infrared luminosity
from the general interstellar radiation field are present in NGC 7552.

\end

\ref{Kawara, K., Nishida, M., and Gregory, b. 1987, ApJ,
321, L35}

\ref{Kent, S. M. 1987, AJ, 93, 816}

\ref{Lester, D. F., Joy, M., Harvey, P. M., Ellis, H. B., Jr.,
and Parmar, P. S. 1987, ApJ, 321, 755}

\ref{Sargent, A. I., Sanders, D. B., and Phillips, T. G.
1989, ApJ, 346, L9}

\ref{Scoville, N. Z., et al. 1989, ApJ, 345,
L25}

\ref{Soifer, B. T., et al. 1986, ApJ, 304, 651}

NGC 7331:
The absolute positional accuracy of these data is estimated
to be $\pm$7$''$; the relative positions of the detectors
are known to $\pm$3\arcsec along a bank, and $\pm$5$''$ between
the banks.

\end